\documentclass[twocolumn,usenatbib]{aastex61}

\usepackage{subfigure}
\usepackage{longtable}
\usepackage{graphicx}
\usepackage{graphics}
\usepackage{keyval}
\usepackage{trig}
\usepackage[american]{babel}
\usepackage{url}
\usepackage{color}
\usepackage{amsmath}
\usepackage{bm}

\def\beq{\begin{equation}}
\def\eeq{\end{equation}}
\def\bey{\begin{eqnarray}}
\def\eey{\end{eqnarray}}

\def\msun{M_\odot}
\def\sun{\odot}
\def\lsim{\mathrel{\raise.3ex\hbox{$<$\kern-.75em\lower1ex\hbox{$\sim$}}}}
\def\gsim{\mathrel{\raise.3ex\hbox{$  $\kern-.75em\lower1ex\hbox{$\sim$}}}}

\def\tz{t_\mathrm{0}}
\def\te{t_\mathrm{E}}
\def\tE{t_\mathrm{E}}
\def\uz{u_\mathrm{0}}

\def\thetae{\theta_\mathrm{E}}
\def\fs{F_\mathrm{S}}
\def\fb{F_\mathrm{B}}
\def\ds{D_\mathrm{S}}
\def\dl{D_\mathrm{L}}

\def\pie{\pi_\mathrm{E}}
\def\piee{\pi_\mathrm{E,E}}
\def\pien{\pi_\mathrm{E,N}}

\def\a0{A_{\mathrm{0}}}

\def\sout{s_\mathrm{out}}

\def\arcsec{^{\prime\prime}}
\def\deltapar{\delta_{\parallel}}
\def\deltaperp{\delta_{\perp}}

\def\c2dof{\chi^2/{\rm d.o.f.}}

\def\muls{\mu_{\mathrm{LS}}}
\def\m606{m_{\rm F606W}}
\def\m814{m_{\rm F814W}}

\def\muas{\mu{\rm as}}
\def\hst{\textit{HST}}

\def\apjl{ApJ}
 
\def\aap{A\&A}      
\def\apjs{ApJS}       

\newcommand\Eq[1]{Eq.~(\ref{#1})}
\newcommand\Fig[1]{Fig.~\ref{#1}}
\newcommand\Tab[1]{Table~\ref{#1}}
\newcommand\Sec[1]{Sec.~\ref{#1}}

\begin{document}

\title{Microlensing constraints on the mass of single stars from HST astrometric measurements}

\author{N. Kains}\affil{Space Telescope Science Institute, 3700 San Martin Drive, Baltimore, MD 21218}\altaffiliation{Based on observations made with the NASA/ESA Hubble Space Telescope, obtained by the Space Telescope Science Institute. STScI is operated by the Association of Universities for Research in Astronomy, Inc., under NASA contract NAS 5-26555, and on observations collected at the European Organisation for Astronomical Research in the Southern Hemisphere under ESO programmes 091.D-0489(A) and 093.D-0522(A).}
\author{A. Calamida}\affil{Space Telescope Science Institute, 3700 San Martin Drive, Baltimore, MD 21218}
\author{K. C. Sahu}\affil{Space Telescope Science Institute, 3700 San Martin Drive, Baltimore, MD 21218} 
\author{S. Casertano}\affil{Space Telescope Science Institute, 3700 San Martin Drive, Baltimore, MD 21218}  
\author{J. Anderson}\affil{Space Telescope Science Institute, 3700 San Martin Drive, Baltimore, MD 21218}  
\author{A. Udalski}\affil{Warsaw University Observatory, Al.~Ujazdowskie~4, 00-478~Warszawa, Poland}  
\author{M. Zoccali} \affil{Pontificia Universidad Cat\'{o}lica de Chile, Instituto de Astrofisica, Av. Vicu\~{n}a Mackenna 4860, Santiago, Chile}\affil{Millennium Institute of Astrophysics, Av. Vicu\~na Mackenna 4860, 782-0436 Macul, Santiago, Chile}
\author{H. Bond}\affil{Space Telescope Science Institute, 3700 San Martin Drive, Baltimore, MD 21218}\affil{Department of Astronomy \& Astrophysics, Pennsylvania State University, University Park, PA 16802, USA}
\author{M. Albrow}\affil{University of Canterbury, Dept. of Physics and Astronomy, Private Bag 4800, 8020 Christchurch, New Zealand}
\author{I. Bond} \affil{Institute of Natural and Mathematical Sciences, Massey University, Auckland 0745, New Zealand}
\author{T. Brown}\affil{Space Telescope Science Institute, 3700 San Martin Drive, Baltimore, MD 21218}   
\author{M. Dominik} \affil{SUPA, School of Physics \& Astronomy, University of St Andrews, North Haugh, St Andrews KY16 9SS, United Kingdom}
\author{C. Fryer} \affil{CCS-2, Los Alamos National Laboratory, Los Alamos, NM 87545, USA}
\author{M. Livio} \affil{Department of Physics and Astronomy, University of Nevada, Las Vegas, 4505 South Maryland Parkway, Las Vegas, NV 89154, USA}
\author{S. Mao}\affil{National Astronomical Observatories, 20A Datun Road, Chinese Academy of Sciences, Beijing 100012, China}
\author{M. Rejkuba}\affil{European Southern Observatory, Karl-Schwarzschild-Strasse 2, D-85748 Garching, Germany}

\begin{abstract}

We report on the first results from a large-scale observing campaign aiming to use astrometric microlensing to detect and place limits on the mass of single objects, including stellar remnants. We used the \textit{Hubble Space Telescope} to monitor stars near the Galactic Center for 3 years, and we measured the brightness and positions of $\sim$2 million stars at each observing epoch. In addition to this, we monitored the same pointings using the VIMOS imager on the Very Large Telescope. The stars we monitored include several bright microlensing events observed from the ground by the OGLE collaboration. In this paper, we present the analysis of our photometric and astrometric measurements for 6 of these events, and derive mass constraints for the lens in each of these. Although these constraints are limited by the photometric precision of ground-based data, and our ability to determine the lens distance, we were able to constrain the size of the Einstein ring radius thanks to our precise astrometric measurements, the first routine measurements of this type from a large-scale observing program. This demonstrates the power of astrometric microlensing as a tool to constrain the masses of stars, stellar remnants, and, in the future, of extrasolar planets, using precise ground- and space-based observations.

\end{abstract}

\keywords{microlensing --- astrometry --- black holes --- exoplanets --- surveys}

\section{Introduction} \label{sec:intro}

When stars and compact objects move within close alignment, both of one another and with respect to an observer, the gravitational deflection of light from the background ``source" object by the ``lens" object leads to the formation of multiple images of the source, an effect known as gravitational lensing. In the case of microlensing, individual images cannot be resolved due their small separation, but the total brightness of the images is larger than the unlensed source, leading to a temporary magnification of the source. This photometric effect has been used extensively, most notably in the search for extrasolar planets \citep[e.g.][]{beaulieu06, gaudi08, kains13b, street16}, as well as the direct mass measurement of several isolated stars \citep[e.g.][]{jiang04, ghosh04}, thanks to second-order effects that can be observed under certain conditions.

In addition to this photometric effect, a gravitational microlensing event produces an astrometric deflection as the event unfolds. This is because the images produced by the lens are not symmetrically distributed, leading to the observed centroid of the source shifting during the event \citep[e.g.][]{dominik00}. Measuring this shift can then allow us to constrain, or measure directly, the lens mass.

Achieving routine measurements of the masses of isolated objects would prove particularly useful in the context of microlensing exoplanet surveys, as they would allow for the planet masses to be tightly constrained. Without mass measurements of the planets' host stars, only the ratio of the planet's mass to that of its host is typically known, with some additional probabilistic constraints derived using Galactic models. Better constraints on planet masses are crucial to our understanding of planet populations, and to our knowledge of exoplanet demographics, particularly for cold, low-mass exoplanets that are best probed with microlensing. Furthermore, mass constraints from microlensing are also an excellent technique to investigate populations of black holes, especially stellar-mass \citep[e.g.][]{lu16} and intermediate-mass black holes \citep{kains16}. Photometric and astrometric microlensing are currently the only known way to probe isolated stellar-mass black holes, with all current information on these objects coming from their effect on companions. 

In this paper, we first recall the main elements of photometric and astrometric microlensing (\Sec{sec:microlensing}), before describing our observations, taken with the \textit{Hubble Space Telescope} (\hst) and the Very Large Telescope (VLT) in \Sec{sec:observations}, and data reductions are discussed in \Sec{sec:reduction}. We present the modelling of our photometric and astrometric data in \Sec{sec:modelling}, and discuss the corresponding constraints they provide on the properties of the lenses, as well as implications for the potential of future space-based microlensing surveys in \Sec{sec:discussion}. Our findings are summarized in \Sec{sec:conclusion}.

\section{Astrometric and photometric microlensing}\label{sec:microlensing}

In this section, we outline the key equations of photometric and astrometric microlensing. For an in-depth discussion of these effects, we refer the reader to \cite{paczynski96} and \cite{dominik00}, respectively.

When an observer, a source at distance $\ds$, and a lens of mass $M$ at a smaller distance $\dl$, move into close enough projected alignment, a microlensing event can occur. The angular separation of the lens and source $\phi$ is usually expressed, as a function of time $t$, in units of the Einstein ring radius, $\thetae$, as $u(t) \equiv u=\phi/\thetae$, where

\begin{equation}\label{eq:thetae}
\thetae = \sqrt{\frac{4GM}{c^2}(\dl^{-1} - \ds^{-1})}\, .
\end{equation}

The lens leads to the production of multiple images of the source, whose total integrated luminosity is larger than that of the unlensed source. This leads to the photometric event, which consists in the magnification of the source by a factor \citep[e.g.][]{paczynski86}

\begin{equation}\label{eq:muu}
\mu(u) = \frac{u^2 + 2}{u\sqrt{u^2+4}}\, .
\end{equation}

\begin{figure}
  \centering
  \includegraphics[width=8cm, angle=0]{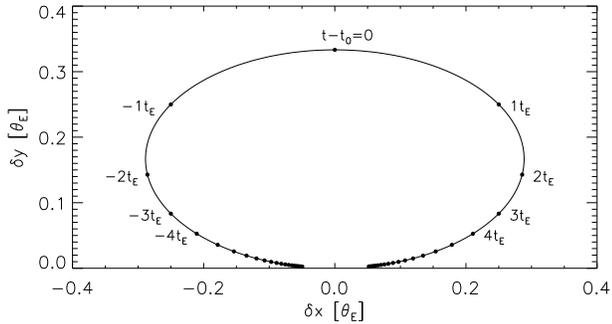}
  \caption{The elliptical trajectory of a source star caused by astrometric microlensing by a lens with an Einstein ring radius of size $\thetae$. Successive filled circles show the source position in increments of $\te$. \label{fig:shift_pos}}
\end{figure}

Because the images of the source are not symmetrically distributed, the apparent centroid of the source also shifts during the event, corresponding to the astrometric part of the microlensing event. The shift $\delta(u)$ can be expressed as \citep{hog95}

\begin{equation}\label{eq:deltau}
\delta(u)=\frac{u}{u^2+2}\thetae\, ,
\end{equation}

\noindent
and points away from the lens, from the observer's standpoint. The parallel and perpendicular components of the displacement, relative to the source-lens relative motion, can be expressed (e.g. \citealt{dominik00}) as

\begin{gather}
\deltapar=\frac{p}{\uz^2 + p^2 + 2}\thetae \nonumber \\
\deltaperp=\frac{\uz}{\uz^2 + p^2 + 2}\thetae \label{eq:deltacomp}\, ,
\end{gather}

\noindent
where $\uz$ is the minimum source-lens angular separation, in units of $\thetae$, also referred to as the impact parameter. This occurs at time $\tz$, and

\begin{equation}\label{eq:p}
p\equiv p(t)=\frac{t-\tz}{\te}\, ,
\end{equation}

\noindent
where $t$ is the time, and $\te$ is the Einstein timescale, with $\te=\thetae/\muls$, where $\muls$ is the source-lens relative motion. Note that \Eq{eq:deltacomp} assumes a rectilinear uniform source-lens relative motion, and is independent of the observational point spread function (PSF). As the source moves relative to the lens, the components of the astrometric shift lead to a characteristic elliptical motion of the source's centroid, as shown in \Fig{fig:shift_pos}. These ellipses have eccentricity $\epsilon=[2/(u_0^2 + 2)]^{1/2}$ \citep{dominik00}.

From \Eq{eq:deltacomp}, measuring the astrometric shift can enable us to measure $\thetae$, which can, via \Eq{eq:thetae}, allow us to make a direct lens mass measurement, provided the lens and source distances can be estimated. This is possible by measuring the effect of annual parallax caused by the Earth's orbit around the Sun, on the photometric light curve of the microlensing event \citep[e.g.][]{dominik98, an02, gould04}. By fitting the components of the parallax vector $\pie$ projected onto the Sky along the the East and North equatorial coordinates, $\piee$ and $\pien$ respectively, we can calculate $\pie$ as

\begin{equation}\label{eq:pie}
\pie = \sqrt{\piee^2+\pien^2} = \frac{\dl^{-1} -\ds^{-1}}{\thetae},
\end{equation}

\noindent
which then allows us to estimate the mass of the lens by reworking \Eq{eq:thetae} as

\begin{equation}\label{eq:mthetae}
M=\frac{\thetae c^2}{4G\pie}\, .
\end{equation}

\section{Observations}\label{sec:observations}

\subsection{OGLE}
Each of the events presented in this paper was initially observed and alerted by the OGLE Early Warning System \citep{udalski03} as part of the OGLE-IV survey \citep{udalski15b}. Observations were taken with the 1.3m Warsaw University Telescope at Las Campanas Observatory, Chile. For full details on the OGLE telescope and CCD setup, as well as observing cadences, see \cite{udalski15b}.

\subsection{\hst}
Our \hst observations were taken in 2012, 2013, and 2014, from mid-March to mid-October, at a cadence of $\sim$ 2 weeks, as well as an initial epoch in October 2011, totalling 192 orbits. This cadence was chosen to provide sufficient time coverage of the astrometric deflection caused by microlensing of a background source by a massive lens, as well as the photometric signature of such events, which would unfold over the course of months. Detecting events cause by massive, non-luminous lenses, such as black holes or neutron stars, was the program's primary science goal, although in this paper we will not discuss such events, but rather ``regular" microlensing events lasting days to weeks that were detected by OGLE, and took place within our \hst pointings.

These observations were taken as part of \hst programs GO-12586, GO-13463, and GO-13057 (PI: K. C. Sahu), using the Advanced Camera for Surveys' Wide Field Channel (ACS/ WFC, hereafter ACS) and the UVIS channel of the Wide Field Camera 3 (WFC3/ UVIS, hereafter WFC3) in parallel. Each epoch consisted of a pair of observations in each of the F606W and F814W filters, with exposure times varying from 350s to 400s. A few short 50s exposures were also taken each season, to allow us to derive a photometric catalogue of bright stars in our field, which are useful for the photometric and astrometric reductions of \hst data. Due to \hst's mid-year orientation flip, the ACS fields were covered the entire year from April to October, whereas the WFC3 pointings were only covered for half of the time each. A summary of the data set is given in \Tab{tab:hstobs}. 

A separate publication (Sahu et al. 2017, in prep.) will discuss the details of this large observing program, and will focus on an in-depth discussion of the techniques and methods developed in the last decade which enabled us to carry out this large program with \hst, and perform the scientific analysis presented in this paper.

\begin{table}
  \vspace{0.5cm}
\begin{center}
  \begin{tabular}{ccccccc}
\hline
Pointing	&RA	&Dec	&F606W	&F814W	 \\
	&(J2000.0)	&(J2000.0)	&Epochs	&Epochs \\
\hline	   
ACS-1	&17:59:00.8	&-29:12:00	&46	&47	\\
ACS-2	&17:59:03.0	&-29:15:20	&45	&46	\\
ACS-3	&17:59:16.2	&-29:11:51	&46	&47	\\
ACS-4	&17:58:47.3	&-29:15:29	&45	&46	\\

WFC3-1	&17:58:41.3	&-29:07:39	&23	&23	\\
WFC3-2	&17:58:43.5	&-29:10:59	&23	&23	\\
WFC3-3	&17:58:56.7	&-29:07:30	&23	&23	\\
WFC3-4	&17:58:28.1	&-29:11:07	&23	&23	\\
WFC3-5	&17:59:20.1    	&-29:16:21	&25	&25	\\
WFC3-6	&17:59:35.5 	&-29:16:13	&25	&25	\\
WFC3-7	&17:59:22.3	&-29:19:41	&24	&24	\\
WFC3-8	&17:59:07.2	&-29:19:50	&24	&24	\\
\hline
\hline
  \end{tabular}
  \caption{Summary of our \hst ACS and WFC3 observations, with the coordinates of each pointing's center, and the number of epochs in each filter. Note that each epoch consists of a pair of observations. \label{tab:hstobs}}
  \end{center}
\end{table}

\subsection{VIMOS}

\begin{figure}
  \centering
  \includegraphics[width=8cm, angle=0]{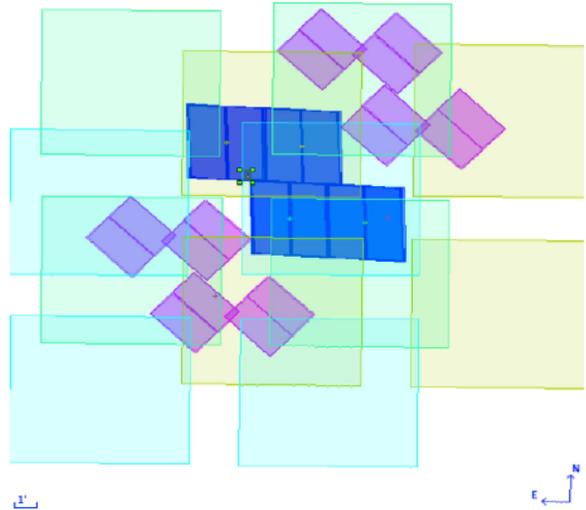}
  \caption{The comparative footprints of our \hst ACS (dark blue trapezoids, center), WFC3 (purple, off-diagonal), and VIMOS (large sets of 4$\times$4 quadrants) pointings. \label{fig:footprint}}
\end{figure}

In addition to our \hst observations, we monitored the same pointings with the VIMOS imager at the VLT at Paranal Observatory, Chile (hereafter referred to as VIMOS data), proposals 091.D-0489(A) and 093.D-0522(A) (PI: M. Zoccali). We covered our \hst footprint with three VIMOS pointings, detailed in \Tab{tab:vimosobs}, in a manner pictured in \Fig{fig:footprint}. VIMOS is a wide-field imager made up of four quadrants, each with a field of view of $7\arcmin$ by $8\arcmin$, separated by a cross-shaped gap 2$\arcmin$ wide. Each quadrant is imaged with a deep-depletion E2V CCD with 2048 x 2440 pixels, and a pixel scale of 0.205$\arcsec$. Further details on this instrument can be found in the instrument reference paper of \cite{lefevre03, hammersley10}. These observations were obtained in 2013 and 2014, from early April to early October, with a higher cadence ($\sim$4 days) than the \hst observations, with the aim of obtaining light curves sampled densely enough to provide constraints on the microlensing parallax (see \Sec{sec:microlensing}) of long microlensing events likely to be caused by massive lenses. Images were obtained mostly in the Bessel-$I$ filter, with a small number of Bessel-$V$ band images also taken during the 2014 season in order to provide colour information on stars in our fields. Exposure times were 30s for most images, as well as a smaller number of 10s exposures in order to be able to construct a reference image with fewer saturated stars. \Tab{tab:vimosobs} summarises the number of images obtained in each band.

\begin{table}
  \vspace{0.5cm}
\begin{center}
  \begin{tabular}{ccccc}
\hline
Pointing	&RA	&Dec	&$I$ 		&$V$  \\
		&(J2000.0)	&(J2000.0) && \\
\hline	   
VIMOS-1	&17:58:42.9	&-29:15:14	&815		&9\\
VIMOS-2	&17:59:11.8	&-29:13:33	&811		&6\\
VIMOS-3	&17:59:18.2	&-29:18:56	&818		&6\\
\hline
  \end{tabular}
  \caption{Summary of our VLT/VIMOS observations, with the coordinates of the centre of each pointing, and the number of epochs in $I$ and $V$. \label{tab:vimosobs}}
  \end{center}
\end{table}

\section{Data reduction}\label{sec:reduction}

\subsection{\hst}
We reduced our \hst\footnote{In the following discussion, we will refer to ``\hst data" instead of ACS and WFC3 data whenever common procedures were used for both data sets, in the interest of simplifying language; we will otherwise refer to data sets by instrument name.} images using the state-of-the-art suite of algorithms by Jay Anderson \citep[e.g.][]{anderson03}. For this work, we used \texttt{flc} images, which have been corrected for charge-transfer efficiency (CTE) losses. As discussed extensively in the literature, CTE losses arise because of detector damages due to cosmic rays, and it is important to correct for their effect when trying to determine the precise positions of stars. 

Stars in each image were detected and measured using the fortran routine \texttt{hst2xym} \citep{anderson06}. We used a standard PSF array for each filter, accounting for spatial variation across the detector; these standard PSF libraries are also provided by \cite{anderson06}. For each star location, the four nearest PSFs are interpolated, to create a PSF that is then used for the local measurement. In addition to this, the time dependence of the PSF is also taken into account by calculating perturbations to the standard PSFs for each image. Accounting for dependence on both the location and time yields PSFs that represent the real stellar profiles accurately, allowing for good flux and position measurements. Finally, we obtained deep photometry using the PSF for each image, with the \texttt{KS2} algorithm developed by Jay Anderson \citep[e.g.][]{anderson00}.

Position measurements from ACS images are also affected by significant geometric distortion. Solutions for the distortion were derived by \cite{anderson06} and are applied to improve the precision of star position measurements.

The photometric measurements were calibrated to the VEGAmag system using the zero-points for the instruments and filters published by STScI \citep[see][for the relevant discussion of the ACS and WFC3 zero-points]{bohlin12}.

\subsubsection{Additional corrections}
We applied additional corrections to both photometric and astrometric measurements from the initial KS2 reduction of our \hst data, in order to correct for systematics and residual trends.

For each target star, we selected reference stars within a 200 pixel radius, each with median magnitude $\xi_{k}$. At each epoch $t_i$, we then calculated the offset between the measured magnitude of each star, $\omega_{k, i}=m_{k, i}-\xi_k$. The photometric offset $\kappa_i$ to be applied at $t_i$ to the target star is then the median of the reference stars' offsets at $t_i$, i.e. $\kappa_i$=med[$\omega_{k, i}$]. For a field of constant stars with well-measured magnitudes, no systematics, and Gaussian errors, $\kappa_i \sim 0$; however, systematics are important, in particular in the ACS data due to the mid-year change in orientation of the telescope, meaning that $\kappa_i$ can be significant. For typical stars in our ACS data, the median photometric offset is around $\pm$2.5  milli-magnitudes, with the sign changing seasonally (i.e. a typical star's brightness is either over- or under-estimated by $\sim$2.5 mmag, depending on the telescope's orientation). Stars in WFC3/UVIS fields, on the other hand, are only observed for half seasons, removing the seasonal variation element; for these, the median offset is typically 1 mmag.

For astrometry, we selected Bulge reference stars within a radius of 200 pixels and with a brightness within 1 magnitude of the target. Bulge stars were identified on the basis on their colour and magnitude, with colour selection given in \Tab{tab:colourpop}, based on the colour-magnitude diagram from observations of the Sagittarius Window Eclipsing Extrasolar Planet Search (SWEEPS) project \citep{sahu06, clarkson08}, the location of which is a subfield of the observations discussed in this paper. For each star, we then derived mean proper motions over the 3 years of observations, and subtracted these from the astrometric time-series to obtain residual astrometric shifts $\beta_{k, i}$. For each epoch $t_i$, we performed a quadratic fit to these residuals for all reference stars, to account for 2-D trends with star position $(x_k, y_k)$, in the form

\begin{equation}\label{eq:2dtrend}
F_i(x_k, y_k)=a_0 + a_1 x_k + a_2 y_k + a_3 x_k^2 + a_4 y_k^2 \, ,
\end{equation}
\noindent

where F is the model and $a_n$ are fitted parameters. We then subtracted the modelled trends from astrometric measurements at each epoch to obtain corrected residuals $\beta'_{k, i}=\beta_{k, i}-F_i(x_k, y_k)$. The median correction for typical ACS stars are large, of the order of $\pm$0.02 pixels, with the sign changing seasonally. For WFC3/UVIS stars, the median correction is much smaller, with a median of the order of 0.002 pixels, and no seasonal variation.

\begin{figure}
  \centering
  \includegraphics[width=8.5cm, angle=0]{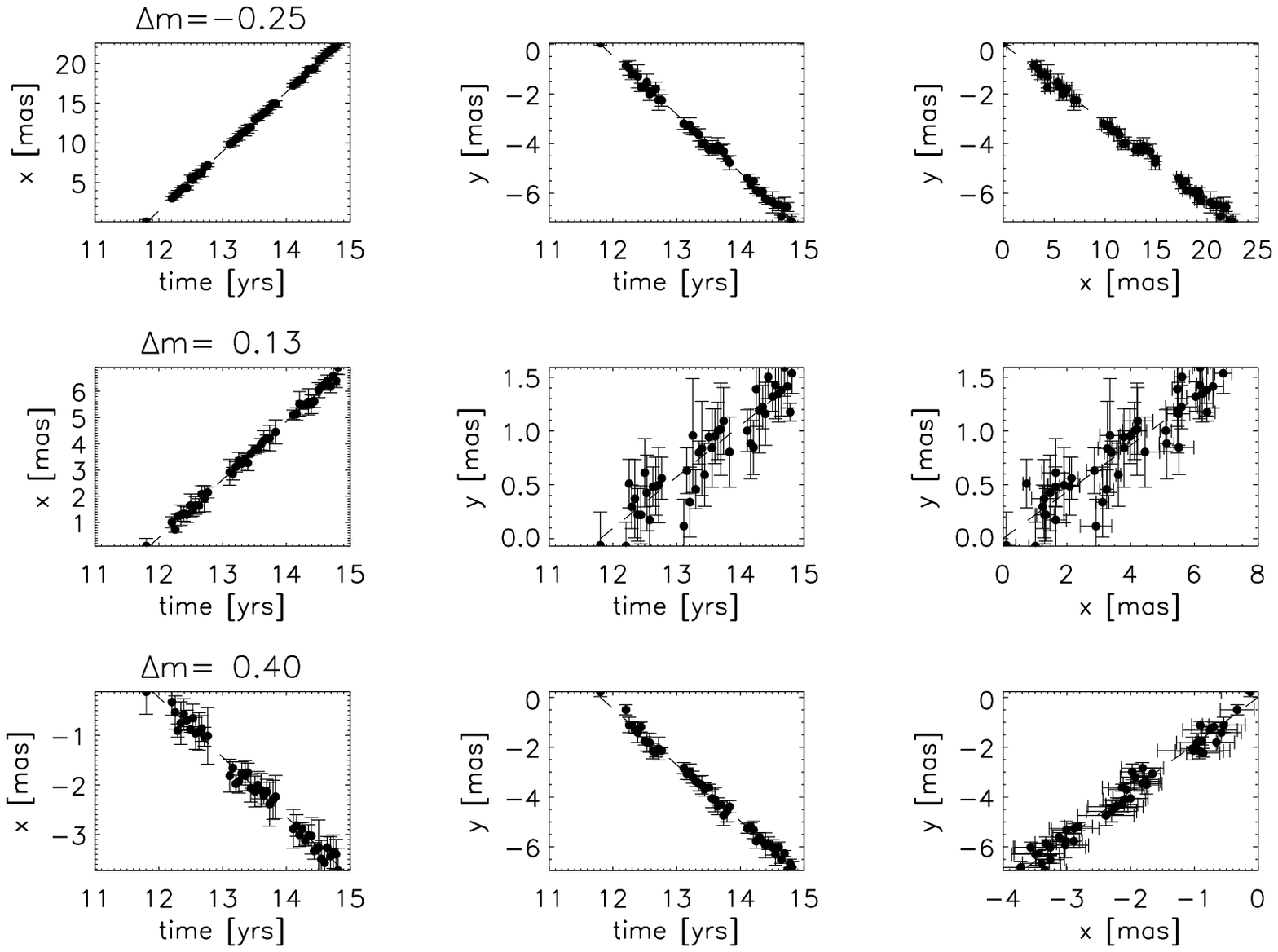}
  \vspace{0.2cm}
  \includegraphics[width=8.5cm, angle=0]{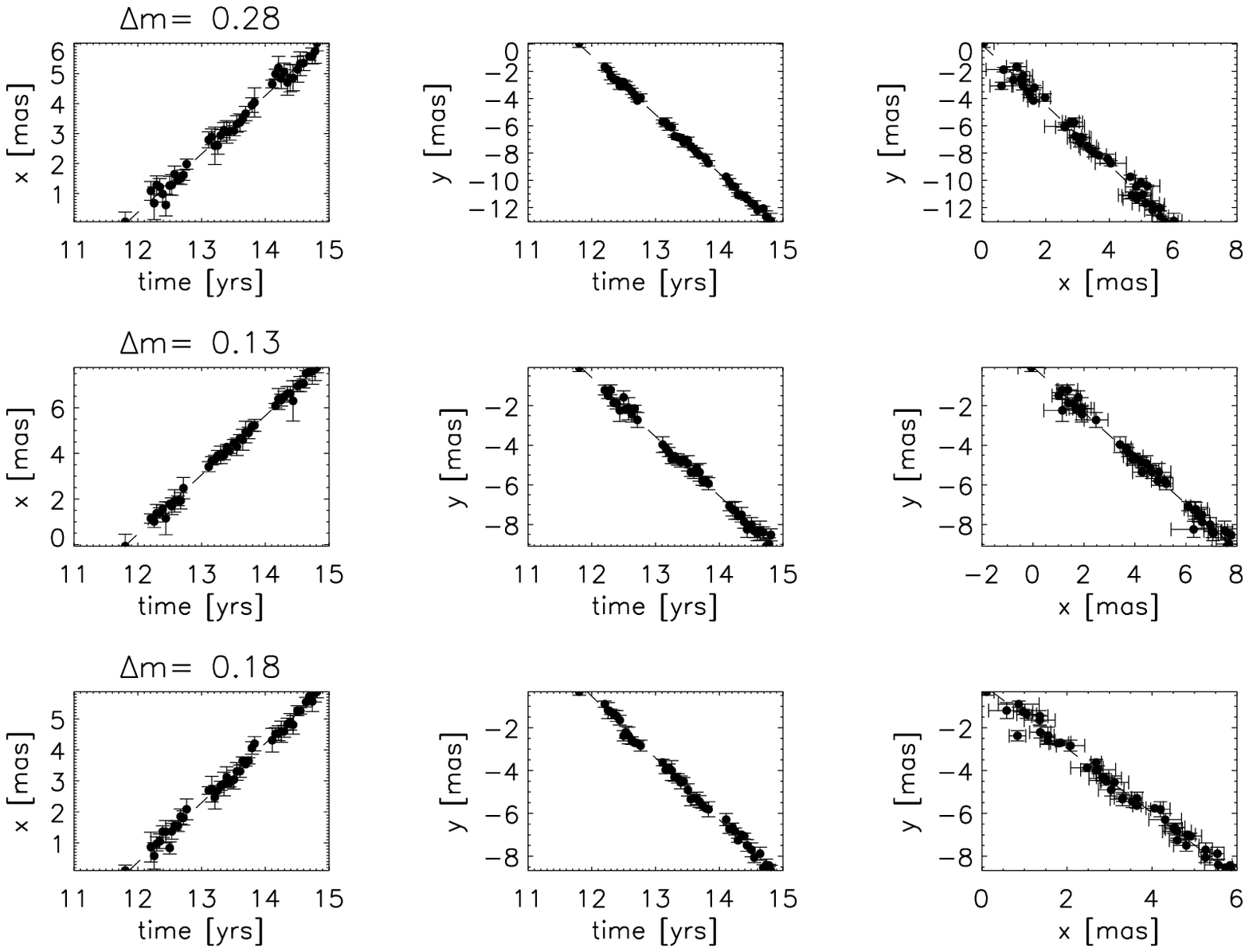}
  \caption{The motions of 6 sample reference stars near OGLE-2013-BLG-0804, plotted as time vs. $x$ and $y$ (left and center columns), and $x$ vs. $y$ (right column). Also plotted as a dashed line is the best linear proper motion model for each star. The magnitude difference of each reference star from the target is given above the left column plot. \label{fig:refstars}}
\end{figure}

Finally, we iterated the entire procedure to obtain final residual astrometric changes not due to proper motion. This was done separately for each dimension, i.e. we fitted residuals in the $x$ direction as a function of $x$ and $y$, and then repeated this for the residuals in the $y$ direction. The resulting time-series are shown in \Sec{sec:fitpar}, while sample reference stars astrometric series are plotted in \Fig{fig:refstars}.

\begin{table}
  \vspace{0.5cm}
\begin{center}
  \begin{tabular}{ccccc}
\hline
$V$ 	&Bulge 		  &a	&b\\
\hline	   
$16.5 < V < 19.5$	& $(V-I) \ge f(V)$ 	&0.0167 		&0.933\\
$20 < V < 21$		& $(V-I) \le f(V)$	&0.06		&-0.04\\
$21 < V < 22$		& $(V-I) \le f(V)$	&0.16		&-2.14\\
$22 < V < 23$		& $(V-I) \le f(V)$	&0.22		&-3.46\\
$23 < V < 25$		& $(V-I) \le f(V)$	&0.367		&-6.83\\
\hline
  \end{tabular}
  \caption{Colour thresholds for the selection of Bulge stars around the SWEEPS field. Bulge stars are those satisfying the conditions in the first two columns, with the function $f(V)=a\,V + b$, and coefficients $a$ and $b$ given in columns 3 and 4. For $19.5<V<20$, it is not possible to separate Bulge and Disk stars using colour alone. Note that these colour selections are only valid in the vicinity of the SWEEPS field, and are not corrected for extinction. \label{tab:colourpop}}
  \end{center}
\end{table}

\subsection{VIMOS}

We reduced the VIMOS images using the difference image analysis (DIA) pipeline \texttt{DanDIA} \citep{bramich13, bramich08}. This pipeline is particularly good at dealing with crowded fields such as the cores of globular clusters \citep[e.g.][]{kains13b, kains12b} and the Galactic Bulge \citep[e.g.][]{kains13a}. Each VIMOS quadrant was reduced separately, resulting in 12 separate independent reduction processes.

We produced a stacked reference using the short (10s) exposures taken at seeing within 10\% of the best-seeing images, in order to minimise the number of saturated stars, which are present in the longer-exposure (30s) images. This resulted in a reference image with the effective exposure times and full-width half-maximum (FWHM) of the PSF listed in \Tab{tab:vimosref}.

\begin{table}
  \vspace{0.5cm}
\begin{center}
  \begin{tabular}{c|cc|cc}
\hline
Pointing	&Exp. [s] 	&FWHM [$\arcsec$]	&Exp. [s] 	&FWHM [$\arcsec$]  \\
\hline
&$I$ &&$V$\\	   
\hline	   
VIMOS-1\\
\hline	   
A.2	&90		&0.61	&20		&0.65	\\
A.3	&100 	&0.56	&30		&0.69	\\
B.1	&210		&0.58	&30		&0.65	\\ 
B.4	&140		&0.58	&30		&0.67	\\
\hline	   
VIMOS-2\\
\hline	   
A.2	&90		&0.59	&30		&0.74	\\
A.3	&140		&0.59	&20		&0.69	\\
B.1	&180		&0.57	&20		&0.69	\\ 
B.4	&130		&0.59	&10		&0.64	\\
\hline	   
VIMOS-3\\
\hline	   
A.2	&60		&0.58	&40		&0.70	\\
A.3	&100		&0.58	&20		&0.58	\\
B.1	&110		&0.59	&40		&0.68	\\ 
B.4	&130		&0.57	&20		&0.57	\\
\hline
  \end{tabular}
  \caption{Characteristics of the $I$ and $V$ reference images for each VIMOS quadrant/ pointing. The total exposure time and effective FWHM in arcseconds are given for each filter. \label{tab:vimosref}}
  \end{center}
\end{table}

We measured the positions and flux of each star in the reference images by extracting a third-degree polynomial empirical PSF from each image, and fitting this PSF to each star. A linear transformation was derived using a triangulation algorithm, in order to register it with the reference. The reference image, convolved with a spatially variable kernel, was subtracted from each image in the series, and resulting difference fluxes were measured at each previously determined star position. Finally, we used the method of \cite{bramich12b} to derive photometric offsets to be applied to each epoch, in order to correct for errors in fitted values of the photometric scale factor. This step was shown by \cite{kains15a} to lead to significant improvements in light curve photometric scatter. Further details of the algorithms used for each step can be found in \cite{bramich08}.

\subsubsection{Astrometry}

\subsubsection{Photometric calibration}

We calibrated the VIMOS photometry using stars in common with OGLE, by matching them using their coordinates. We derived a colour-dependent calibration relation for each field in the form

\begin{equation}\label{eq:vimoscal}
I_{\mathrm{cal}}=I_{\mathrm{ins}} + a + b(V_{\mathrm{ins}} - I_{\mathrm{ins}}),
\end{equation}

\noindent
where $I_{\mathrm{cal}}$ ($V_{\mathrm{cal}}$) is the calibrated magnitude, $I_{\mathrm{ins}}$ ($V_{\mathrm{ins}}$) is the instrumental magnitude measured by the DIA photometry, and $a$ and $b$ are the transformation coefficients. 

\begin{figure}
  \centering
  \includegraphics[width=8.5cm, angle=0]{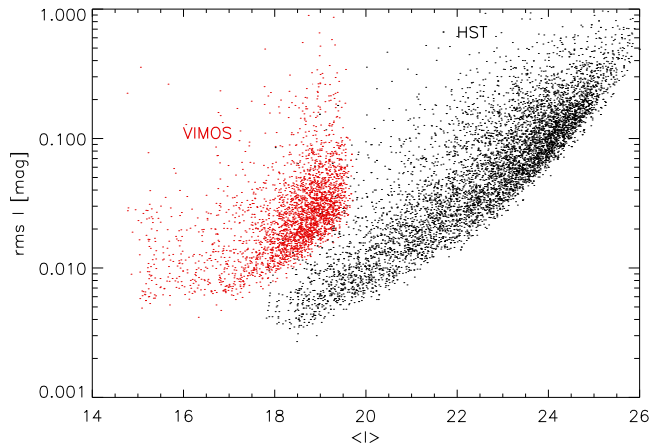}
  \caption{The photometric rms scatter for stars in the \hst (black, right side of the plot) and VIMOS (red, left) data. \label{fig:rms_hst_vimos}}
\end{figure}

\Fig{fig:rms_hst_vimos} shows the root mean square (rms) scatter of the lightcurves of stars in the both our \hst and VIMOS photometry. The range of magnitudes covered overlaps by only $\sim 1.5$mag, with \hst observations saturated at $I\sim18$, and VIMOS observations being no deeper than $I\sim19.5$. The typical photometric rms scatter is less than 0.01 for stars down to $I\sim 20.5$ for \hst observations, whereas that is the case for stars brighter than $I\sim18$ in the VIMOS data.

\subsection{OGLE}

The OGLE observations were reduced with their optimised offline pipeline, which is built from the standard OGLE pipeline \citep{udalski03}, but optimised for use on OGLE-IV data after determining centroid for the source star of each event manually.

\section{Modelling}\label{sec:modelling}

\subsection{Event selection}

We selected OGLE events based on the quality of the \hst astrometric measurements, the value of $\te$, and the event reaching its photometric peak within our \hst observing campaigns, to ensure good coverage of any astrometric signal to yield the best possible mass constraint. Events with a $\te$ much shorter than our observing cadence of $\sim2$ weeks were not analysed, given the small size of their expected astrometric signals, as were events where residuals showed residual ``seasonal" trends with the telescope orientation. The selected events are listed in \Tab{tab:eventlist}. Of the events satisfying these conditions, one event, OGLE-2014-BLG-0442, was also rejected because the photometry showed clear deviations from point source-point lens (PSPL) microlensing. One event, OGLE-2014-BLG-1045, was rejected due to large residual seasonal trends which we were unable to correct for, and one event was saturated due to close proximity to a 15$^{\mathrm{th}}$ magnitude star. The remaining six events are the bright single-lens events in our program that could be followed from the ground and space simultaneously. We also detected a number of events with faint sources that could only be observed from space; these events will be discussed in a separate publication.

\begin{table*}
  \vspace{0.5cm}
\begin{center}
  \begin{tabular}{ccccc}
\hline
OGLE event   &RA 	&Dec 	&\hst Pointing 	&VIMOS Pointing\\
	&(J2000.0) &(J2000.0) &(Tab. \ref{tab:hstobs}) 	&(Tab. \ref{tab:vimosobs})\\
\hline	   
2012-BLG-0645 &17:58:58.8		&-29:13:44	&ACS-2	&$-$\\
2013-BLG-0182 &17:58:41.2		&-29:07:25	&WFC3-1	&VIMOS-1 A.3$^*$\\
2013-BLG-0547 &17:59:00.2		&-29:10:32	&ACS-1	&VIMOS-2 A.3$^*$\\
2013-BLG-0804 &17:58:41.1		&-29:15:23	&ACS-4	&VIMOS-3 A.3$^*$\\
2013-BLG-1547 &17:59:16.3		&-29:13:15	&ACS-3	&VIMOS-1 A.3\\
2014-BLG-0117 &17:59:07.0		&-29:11:38	&ACS-1	&$-$\\
\hline
  \end{tabular}
  \caption{Single-lens OGLE events in our \hst observations, with RA and Dec coordinates from our \hst reference image, and the \hst pointing within which the event is located (see \Tab{tab:hstobs}). The VIMOS pointing (\Tab{tab:vimosobs}) where the event is covered is also given, with an asterisk when a VIMOS light curve could be extracted. \label{tab:eventlist}}
  \end{center}
\end{table*}

We also note that only three events have VIMOS photometry: since VIMOS observations were taken in 2013 and 2014 only, there is no data for OGLE-2012-BLG-0645. OGLE-2014-BLG-0117 occurred during the gap between 2013 and 2014 observations. OGLE-2013-BLG-1547 could not be found in the VIMOS data, possibly because of its position near the edge of the image, where photometric scatter is larger, and its location near two bright stars, affected our ability to detect this event. Lastly, OGLE-2013-BLG-0182 does have a VIMOS light curve, but only covering the very end of the event.

\subsection{Photometry}

For each event, we first fitted the available light curves, comprising \hst, VIMOS, and OGLE data. We used a Markov Chain - Monte Carlo (MCMC) algorithm to fit the simple PSPL parameters, $\tz$, $\te$, and $\uz$, as well as parallax parameters $\piee$ and $\pien$ (see \Eq{eq:pie}). The latter allow us to fit for the effect of the annual parallax due to the Earth's orbit around the Sun, for which we used the geocentric formalism \citep{dominik98, gould04}. The advantage of this approach is that a good estimate for the parameters $\tz$, $\te$, and $\uz$ can be obtained from a simple PSPL fit that does not include parallax. We also fitted the source and blend flux parameters, $\fs^i$ and $\fb^i$, for each telescope and filter combination, such that the model flux at time $t$ and site $i$ is given by 

\begin{equation}\label{eq:fsfb}
F^i(t)=\fs^i \mu(t) + \fb^i .
\end{equation}

\noindent
These flux parameters are fitted by linear regression for each set of PSPL parameters explored by the MCMC, and allow to account for blended flux detected that is not being lensed. This produced a first fit of the PSPL parameters, along with associated uncertainties from the posterior distribution of each parameter.

\subsection{Astrometry}

\begin{figure}
  \centering
  \includegraphics[width=8.5cm, angle=0]{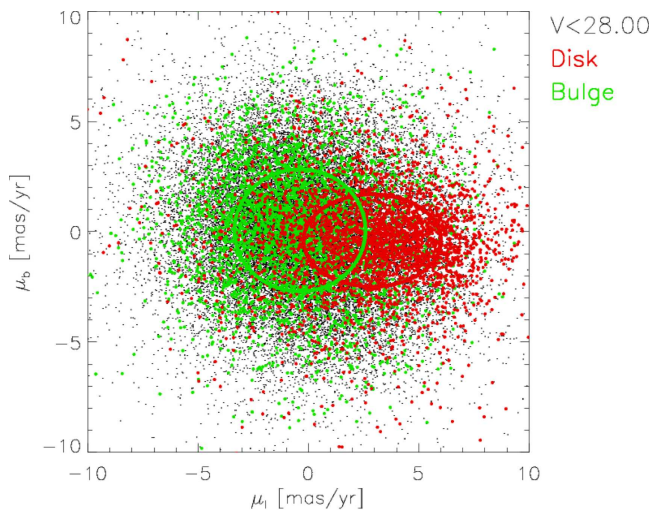}
  \caption{The components of the proper motions of Bulge (red) and Disk (green) stars along the $l$ and $b$ directions. Disk stars occupy a smaller region of the plane, as indicated by the ellipses containing 68.3\% of each population. For easier visualisation, similar numbers of Disk and Bulge stars are plotted, but Disk contamination of the Bulge stars selected by color (\Tab{tab:colourpop}) is only $\sim$4\% at the 3-$\sigma$ level. \label{fig:pm}}
\end{figure}

As detailed in \Sec{sec:microlensing}, an astrometric trajectory is described by the parameters $\tz, \uz, \te$, as well as $\thetae$. In order to describe the elliptical motion fully, an additional parameter $\alpha$ is needed, corresponding to the angle of relative source-lens motion in the plane of the Sky. In addition, we fit two parameters $x_0$ and $y_0$, which correspond to the arbitrary baseline reference position of the source at $t\ll\tz$, when no astrometric deflection due to microlensing is present. Since $\tz, \uz$, and $\te$ are well constrained by the photometry, the posteriors obtained from MCMC fits to the light curves can be used to set priors on these parameters for the subsequent astrometric fits. Prior information on $\alpha$ can also be obtained by considering the proper motion of the source, and the proper motion distribution of Disk stars. As shown in \Fig{fig:pm}, Disk stars lie in a sub-region of the ($\mu_l, \mu_b$) plane, where $\mu_l$ and $\mu_b$ are the two components of the proper motions along the $l$ and $b$ directions, respectively. To obtain this distribution, we first measured the proper motions of stars down to $V\sim$28 mag, and then overplotted Disk and Bulge stars selected according to their colour and magnitude. We then derived median values for the proper motion of each population, along each direction, and 1-$\sigma$ error bars corresponding to the limit of the $68.3\%$ confidence interval. We find values of $(\langle\mu_l\rangle, \langle\mu_b\rangle)=(-0.35, 0.06)$ mas yr$^{-1}$ and $(\sigma_l, \sigma_b)=(2.97, 2.77)$ mas yr$^{-1}$ for Bulge stars, while for Disk stars, we find $(\langle\mu_l\rangle, \langle\mu_b\rangle)=(2.89, -0.40)$ mas yr$^{-1}$ and $(\sigma_l, \sigma_b)=(3.12, 2.09)$ mas yr$^{-1}$. This is in good agreements with values found by \cite{clarkson08} for the SWEEPS field.

For stars with parallax measurements yielding a lens distance corresponding to a Disk star, one can therefore derive a probability distribution for the lens proper motion components, and, using the measurements of the source proper motion, transform this into a probability distribution for the relative angle of motion $\alpha$, which can then be used as a prior. On the other hand, for stars with no parallax measurements, the probability that the lens as well as the source is a Bulge star is high, and we therefore keep the prior on $\alpha$ uniform for these events. Only $\thetae$, $x_0$, and $y_0$ are left as parameters with uniform (non-informative) priors. Finally, we note that although parallax does have a small effect on the astrometry, it is smaller than what can be detected with our data, and we therefore ignore this by not re-fitting $\pie$ and $\pien$ as part of the astrometric model, even when it is well constrained by the photometry. 

\subsection{Final parameters}\label{sec:fitpar}

\begin{figure*}
  \centering
  \includegraphics[width=8.5cm, angle=0]{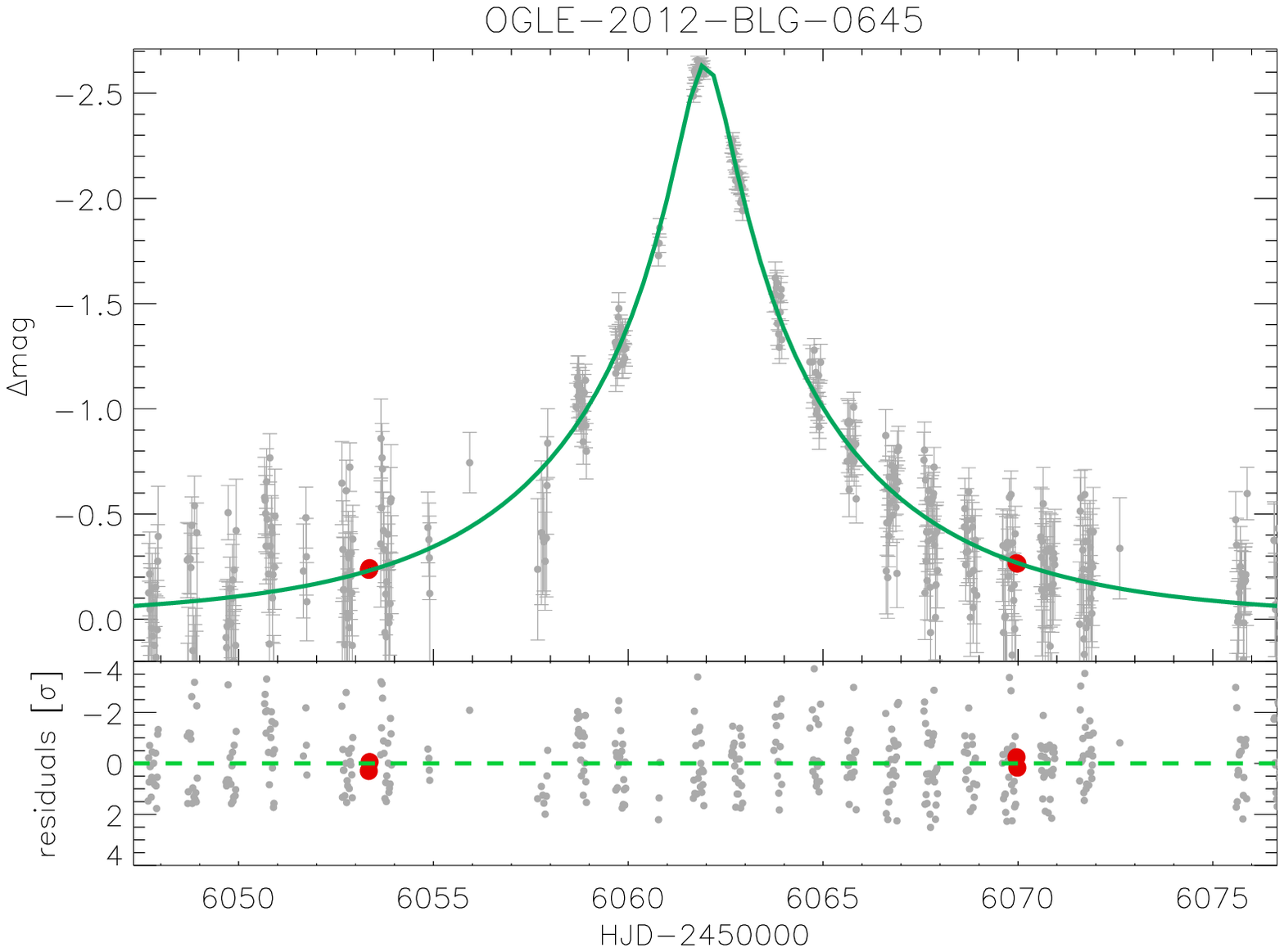}
   \vspace{0.2cm}
  \includegraphics[width=8.5cm, angle=0]{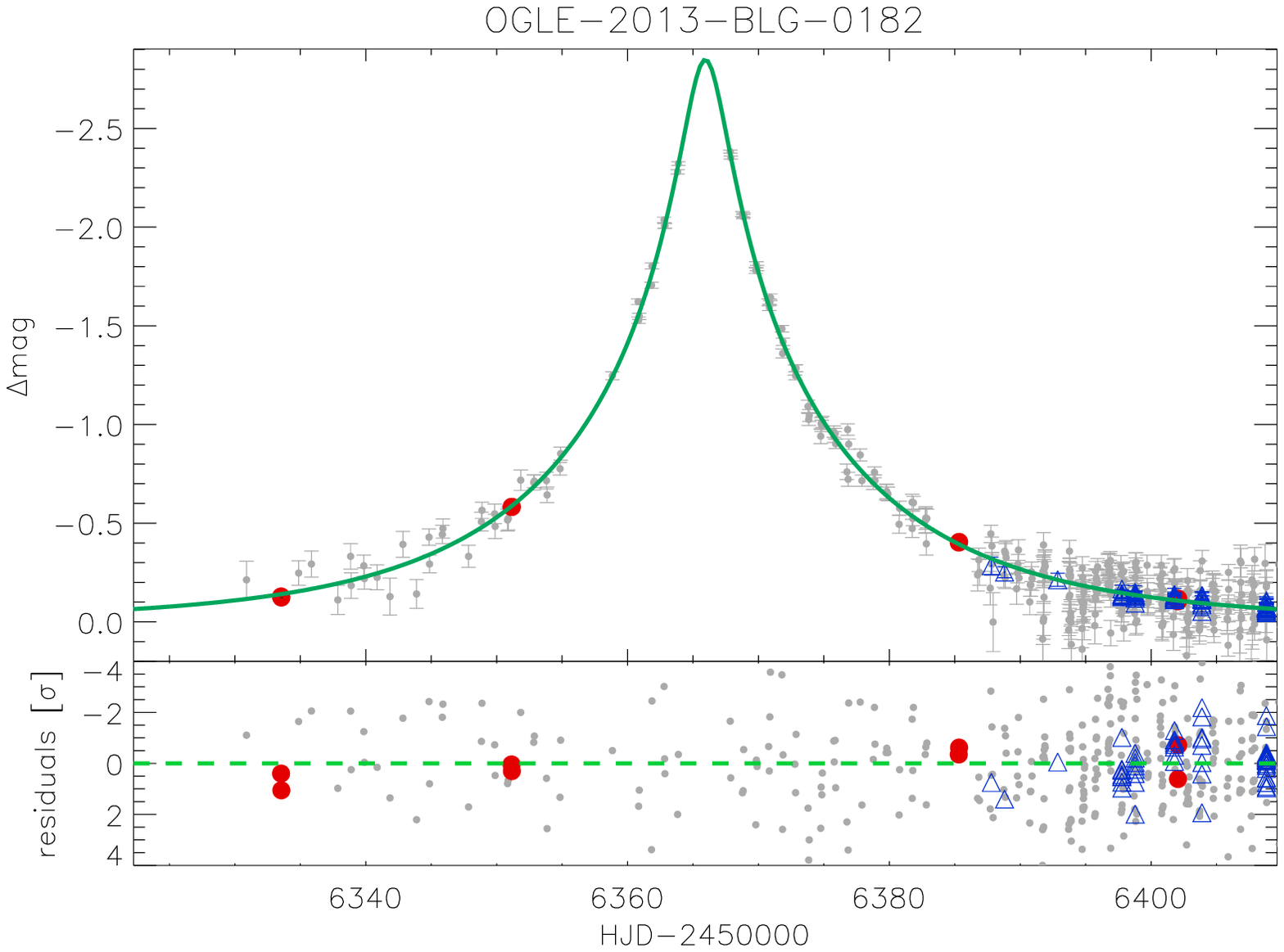}
   \vspace{0.2cm}
  \includegraphics[width=8.5cm, angle=0]{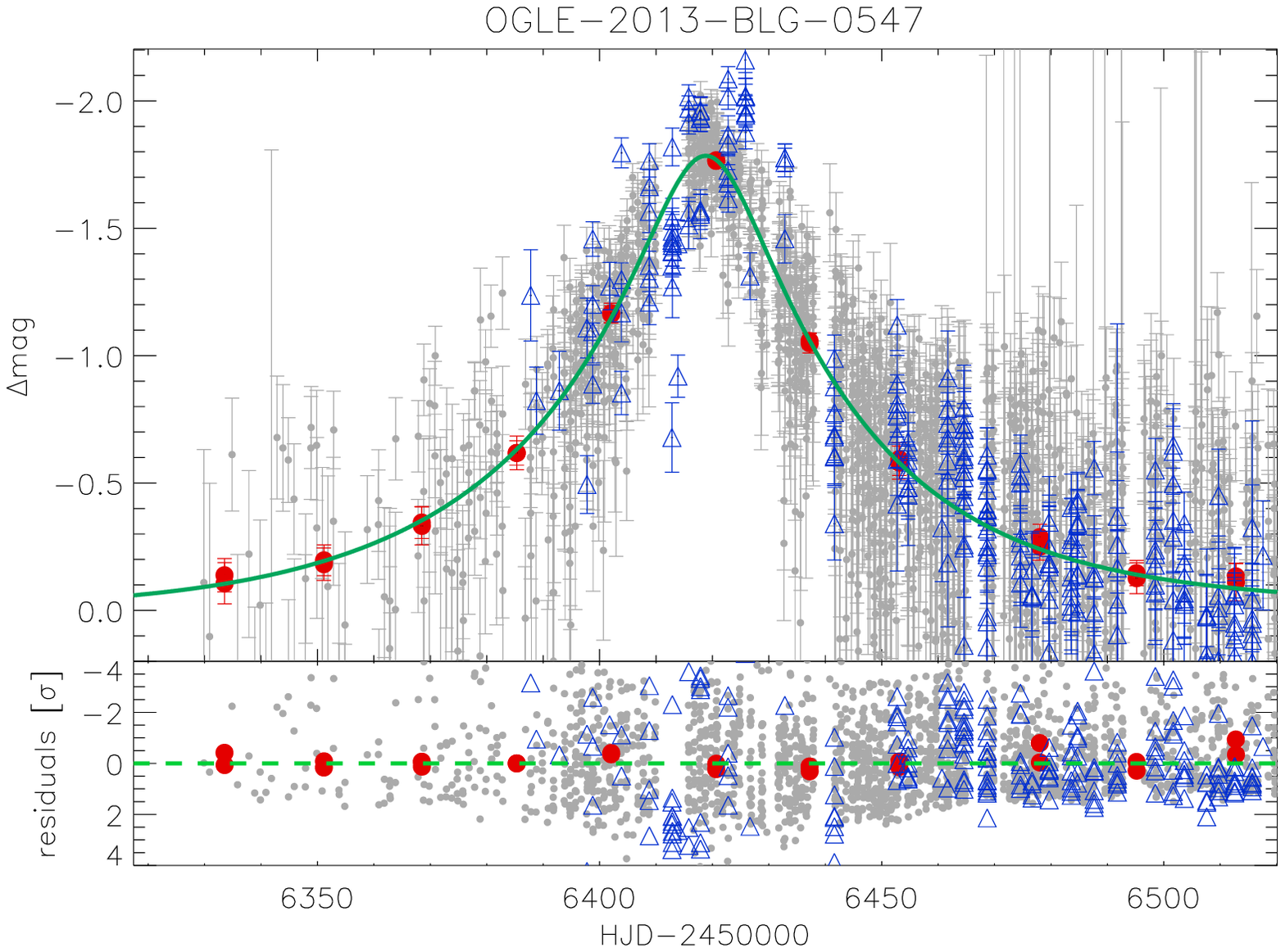}
    \vspace{0.2cm}
  \includegraphics[width=8.5cm, angle=0]{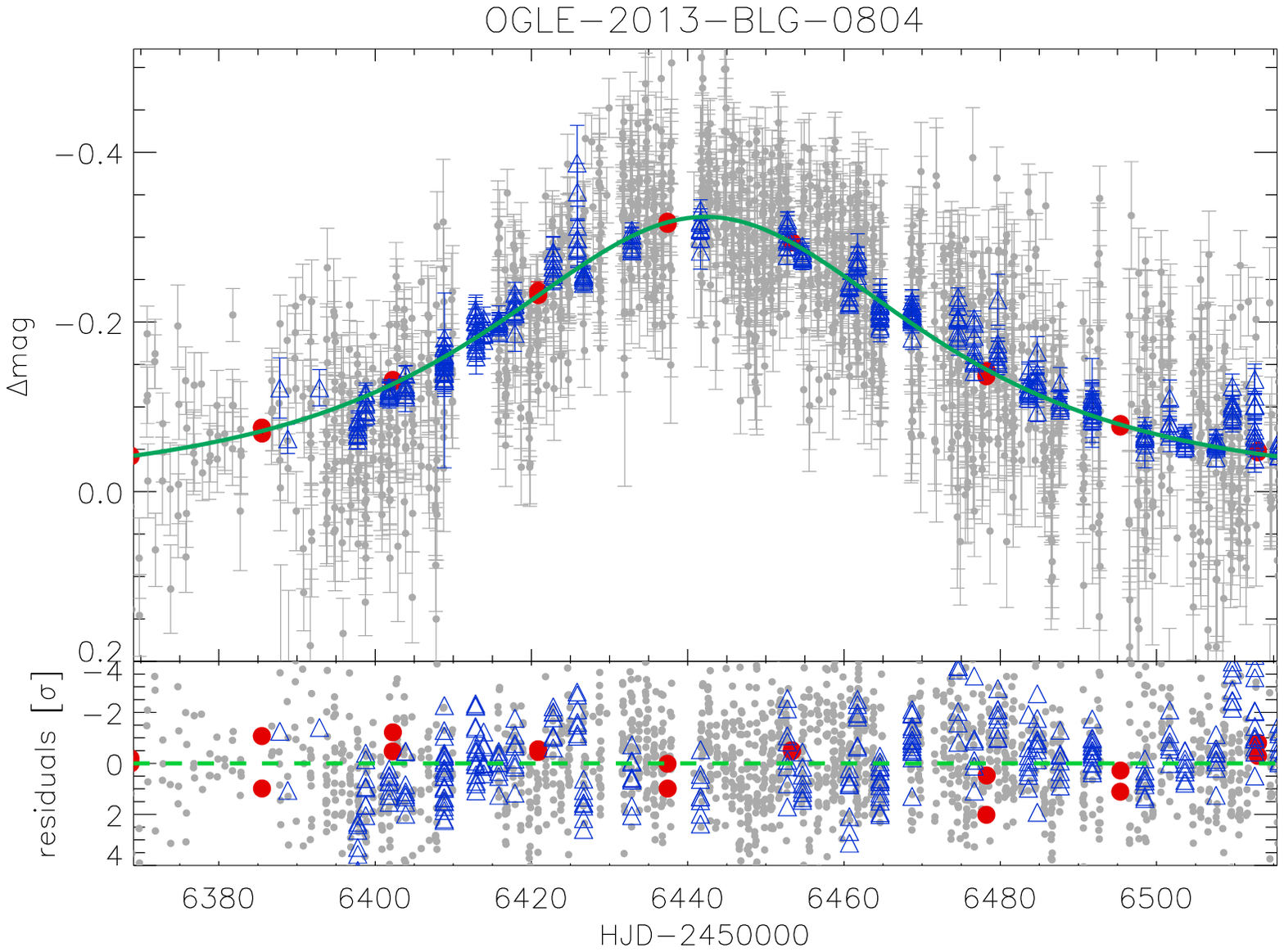}
    \vspace{0.2cm}
  \includegraphics[width=8.5cm, angle=0]{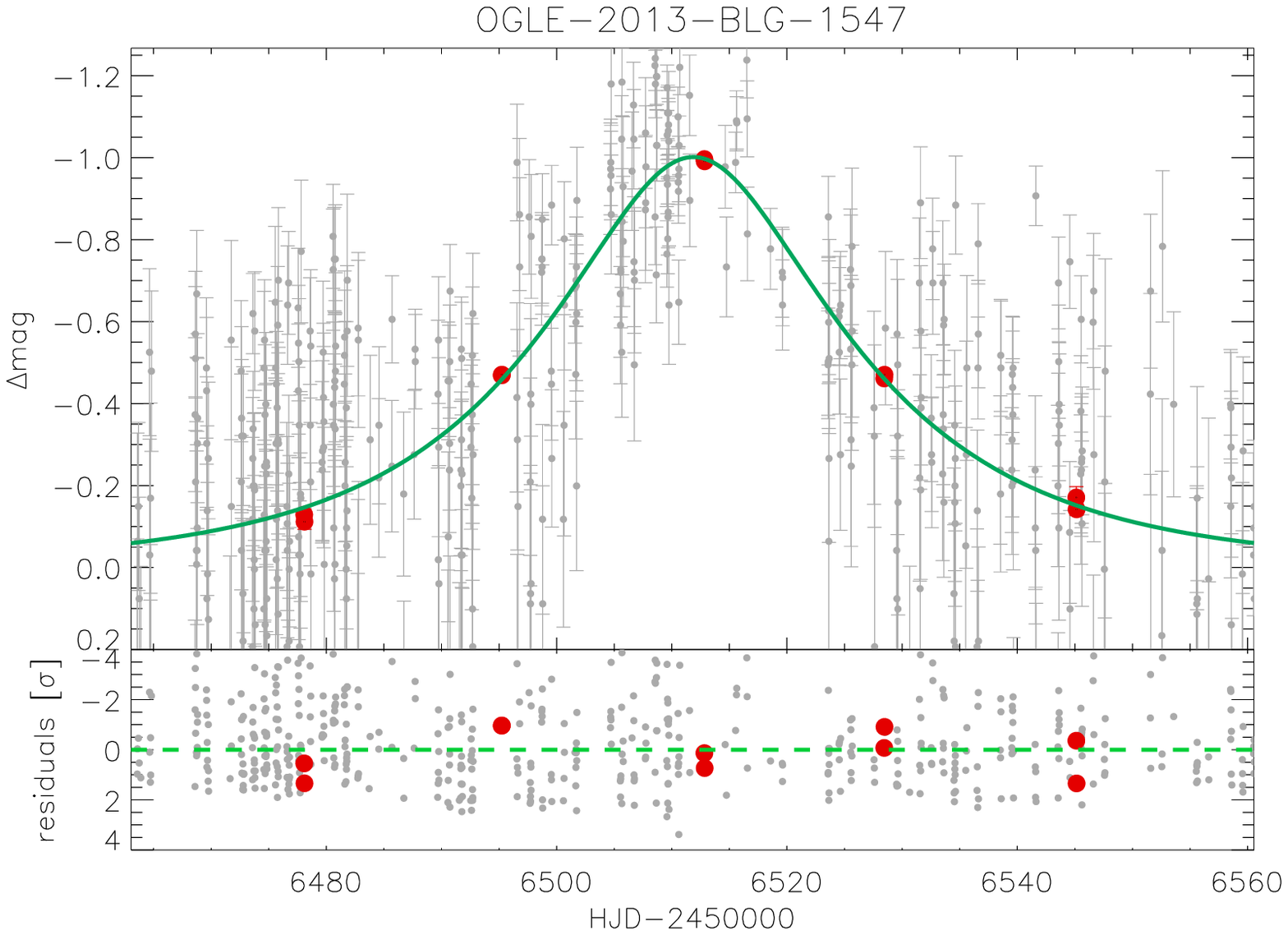}
    \vspace{0.2cm}
  \includegraphics[width=8.5cm, angle=0]{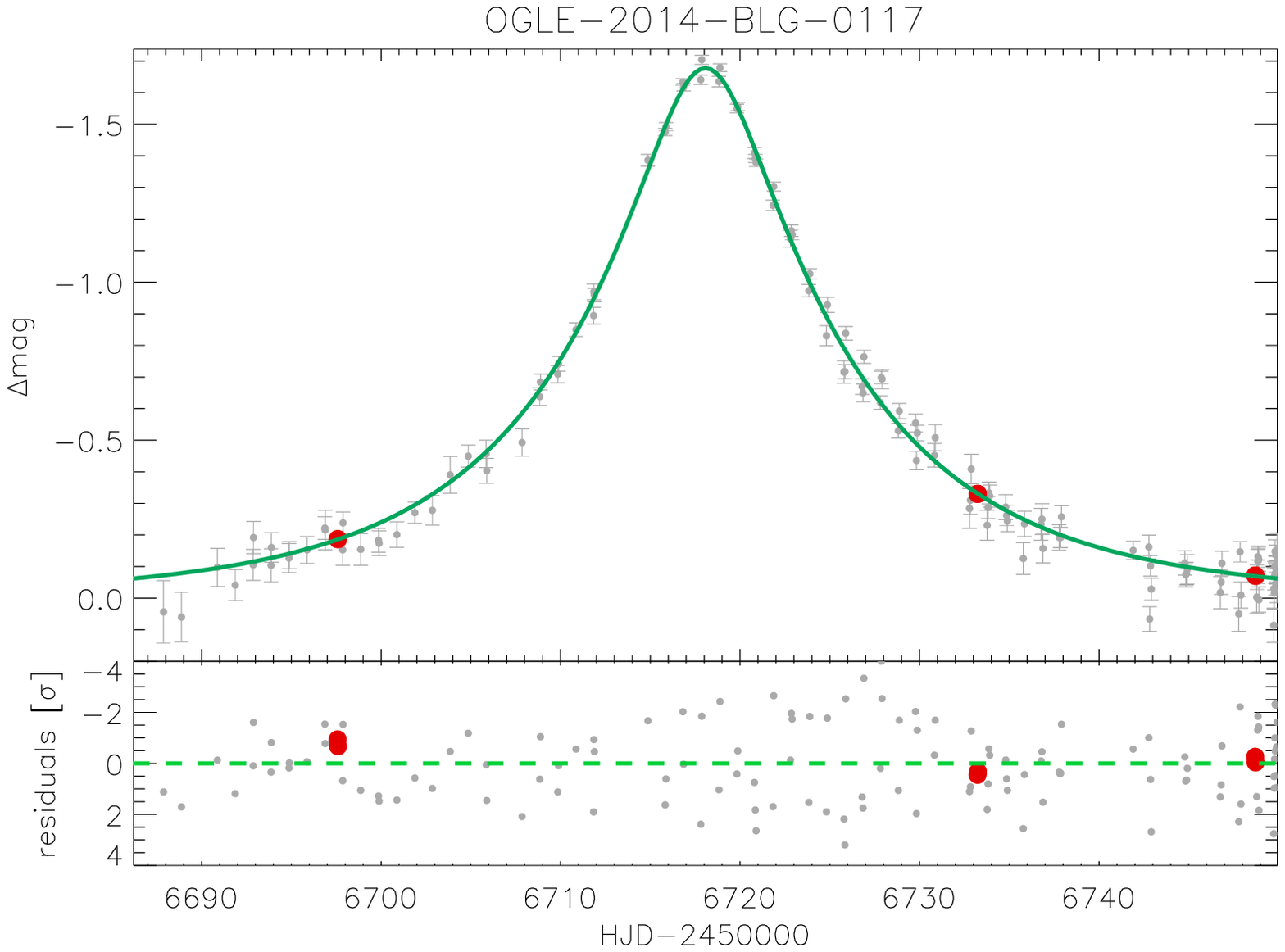}
  \caption{The observations and best-fit model for the photometry of each of our six events. OGLE data are plotted as small grey filled circles, \hst data as large red filled circles, and VIMOS data as blue open triangles; all data are plotted with 1-$\sigma$ error bars.  For each event, the best-fit model light curve is plotted as a green solid curve. \label{fig:photfit}}
\end{figure*}

\begin{figure*}
  \centering
  \includegraphics[width=8.5cm, angle=0]{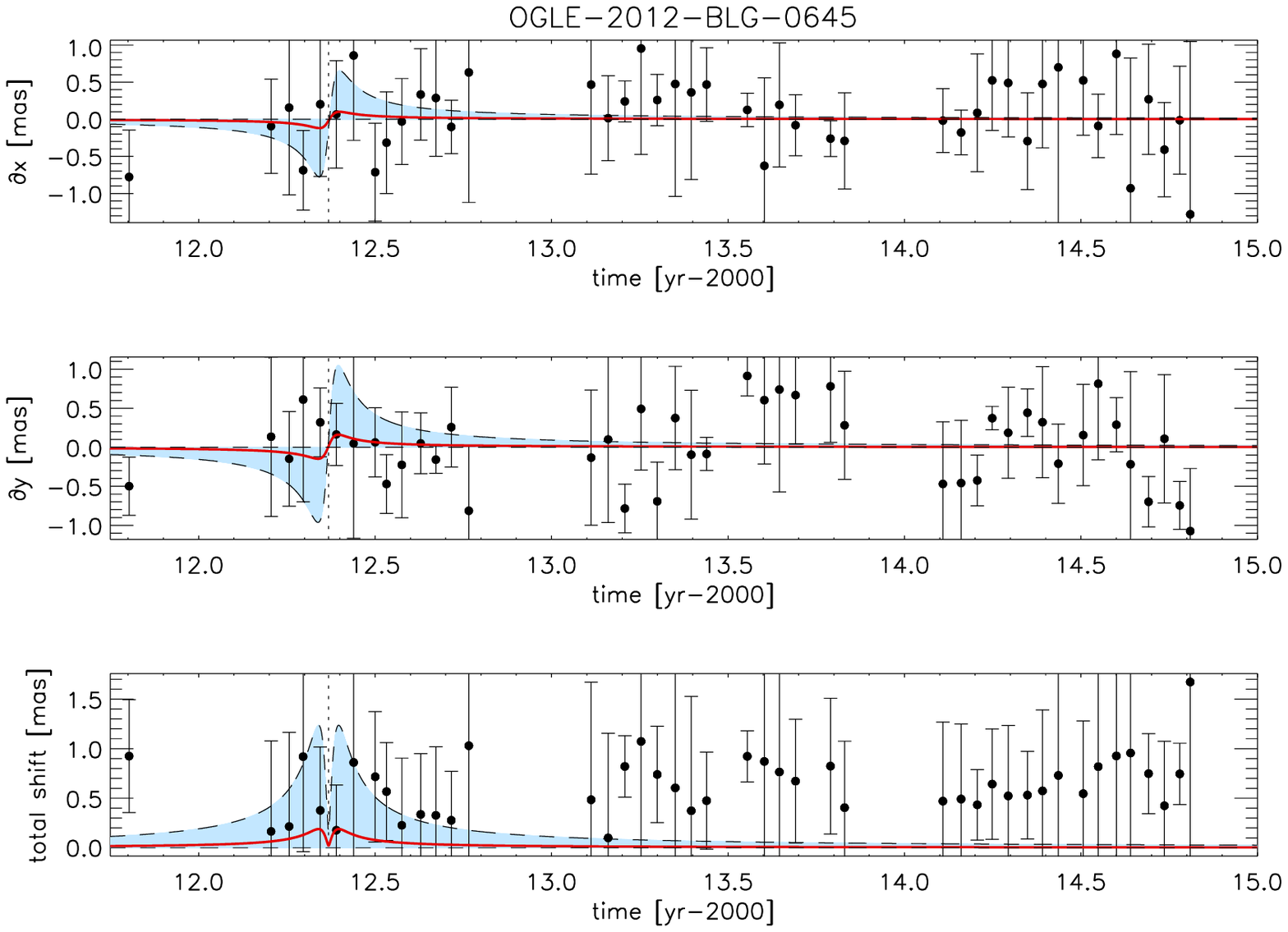}
   \vspace{0.3cm}
  \includegraphics[width=8.5cm, angle=0]{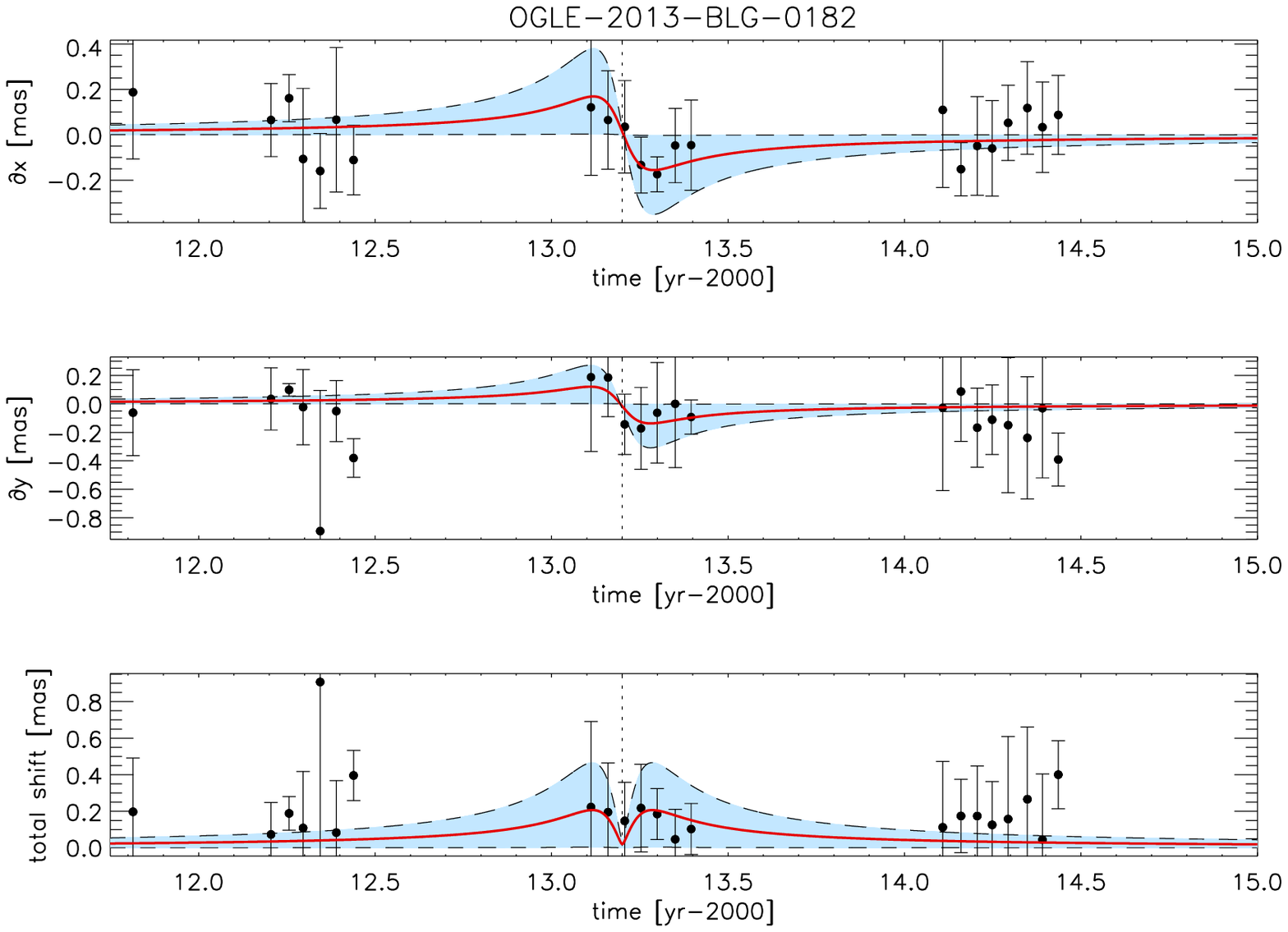}
   \vspace{0.3cm}
  \includegraphics[width=8.5cm, angle=0]{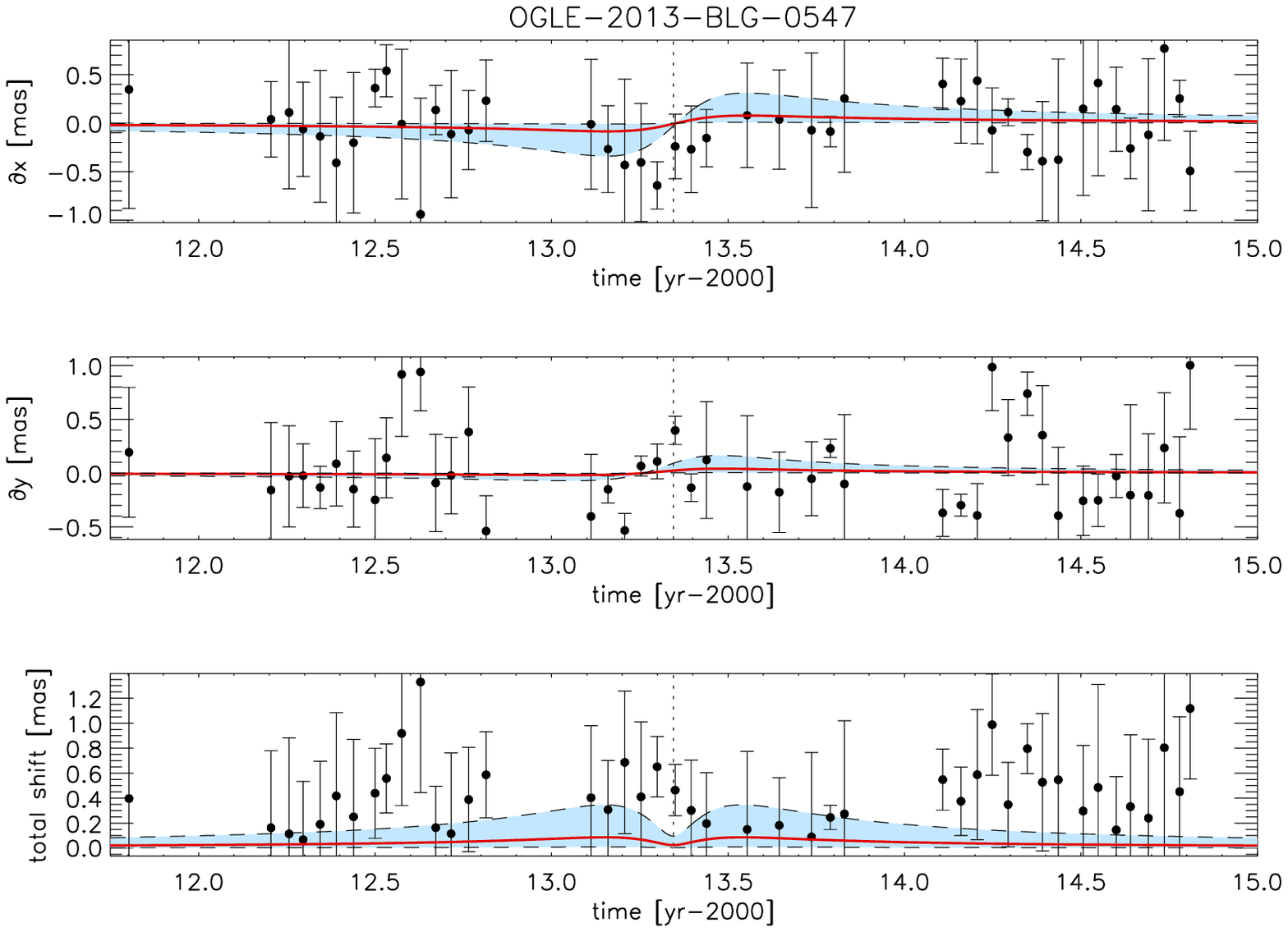}
    \vspace{0.3cm}
  \includegraphics[width=8.5cm, angle=0]{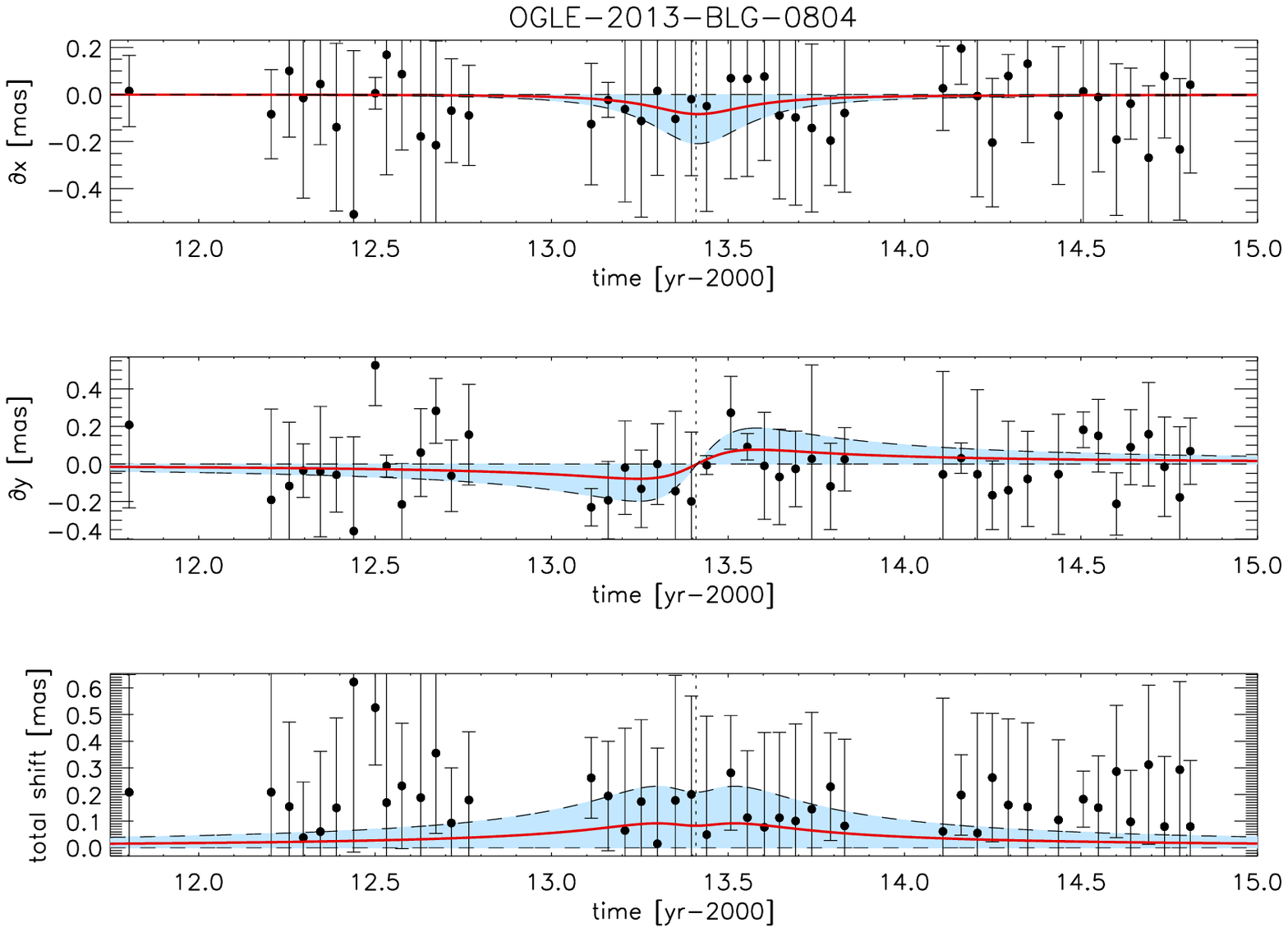}
    \vspace{0.3cm}
  \includegraphics[width=8.5cm, angle=0]{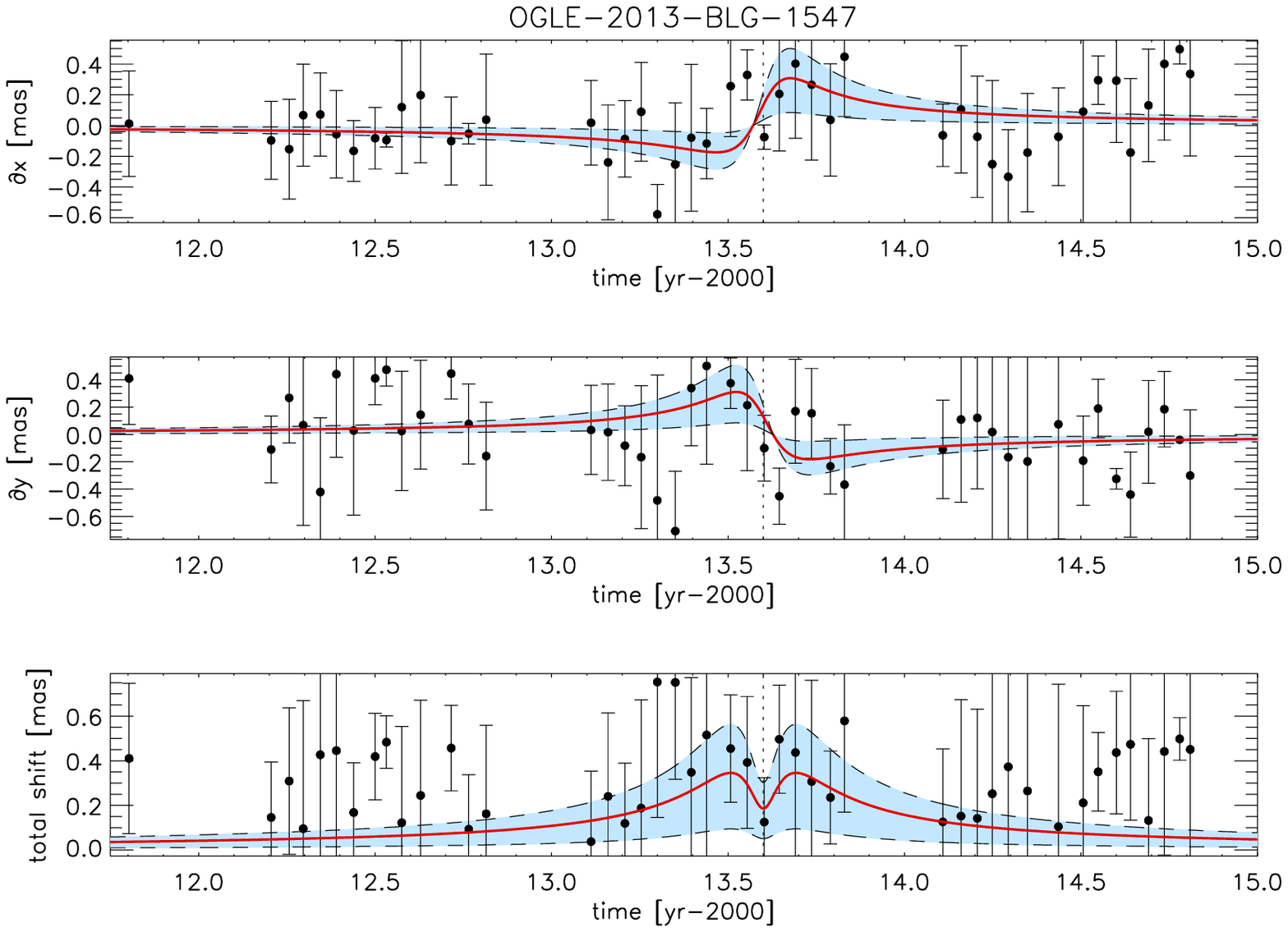}
    \vspace{0.3cm}
  \includegraphics[width=8.5cm, angle=0]{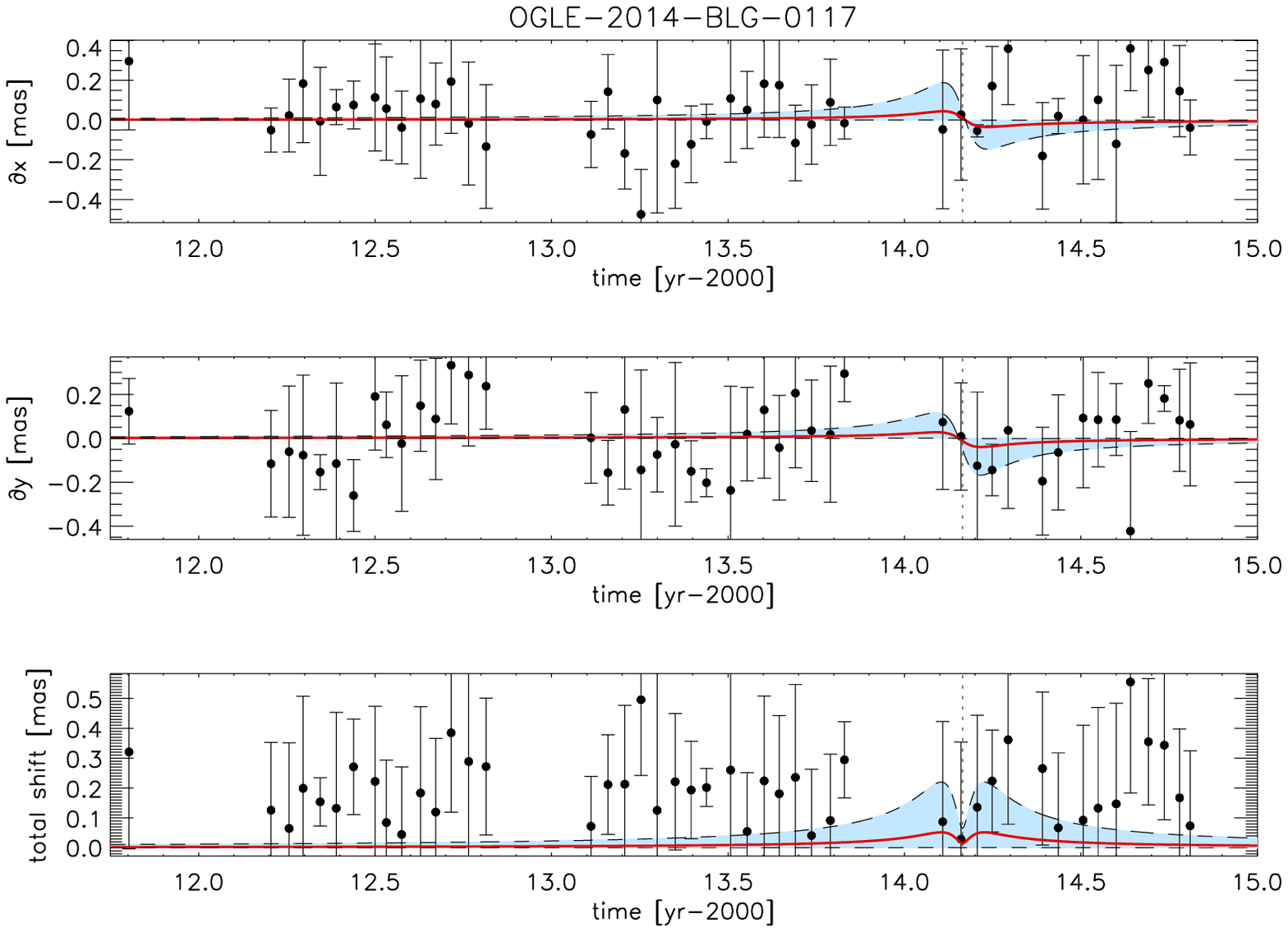}
  \caption{The observations and best-fit model for the astrometric measurements of each of our six events, after subtraction of the mean rectilinear proper motion, along the $x$ and $y$ axes (top and middle panels for each event, respectively), and the total shift $\sqrt{x^2+y^2}$ (bottom panel for each event), as a function of time. For each event, astrometric measurements are plotted as black filled circles with 1-$\sigma$ error bars, and the best-fit model is plotted as a red solid curve, while the blue shaded areas, delimited by dashed lines, show the 99.7\% confidence intervals. Note that the time axis here is in years, rather than days, for easier visualisation of the multi-year astrometric curve. \label{fig:ast1d}}
\end{figure*}

\begin{figure*}
  \centering
  \includegraphics[width=5.5cm, angle=0]{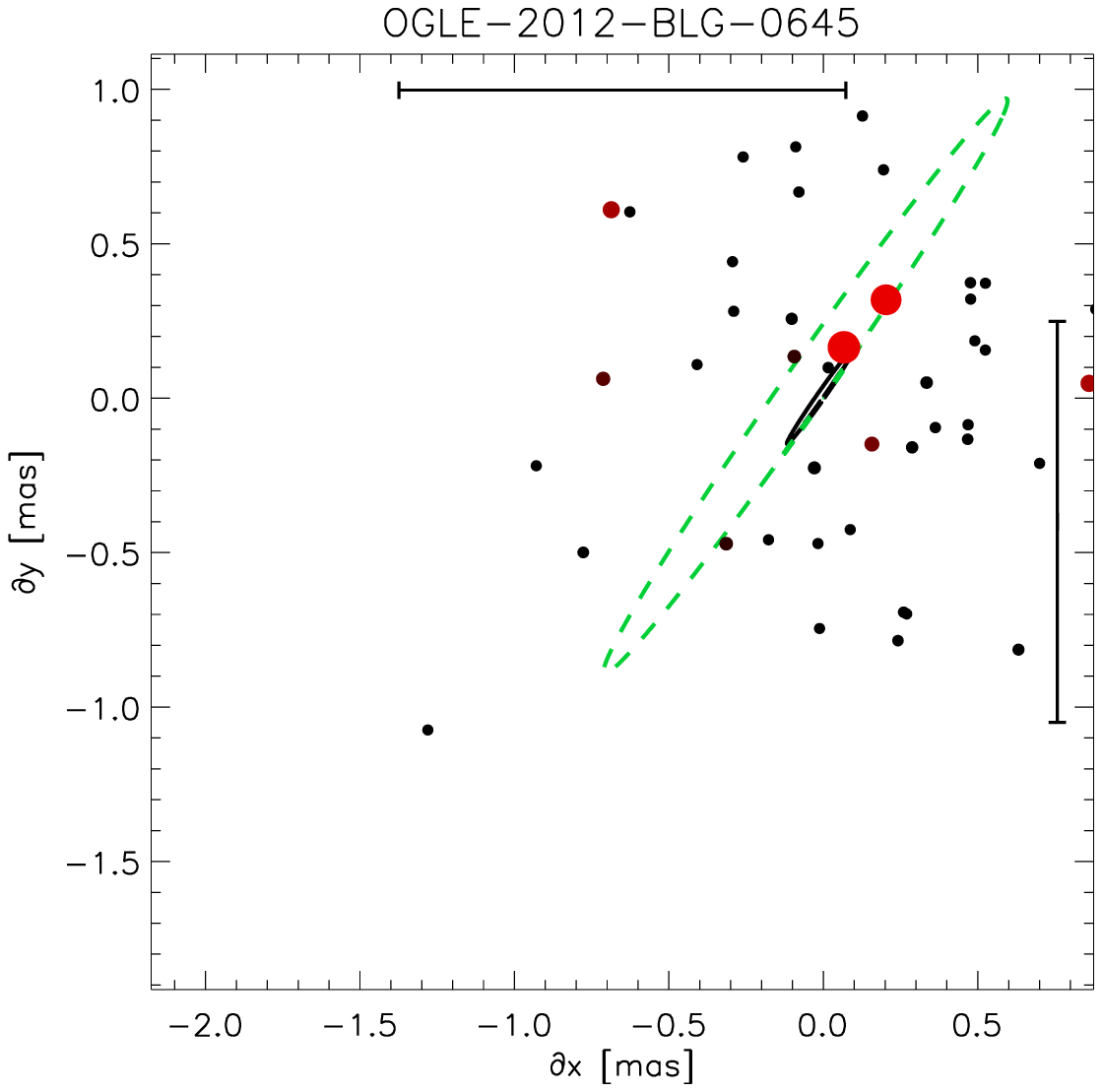}
   \hspace{0.5cm}
    \vspace{0.5cm}   
  \includegraphics[width=4cm, angle=0]{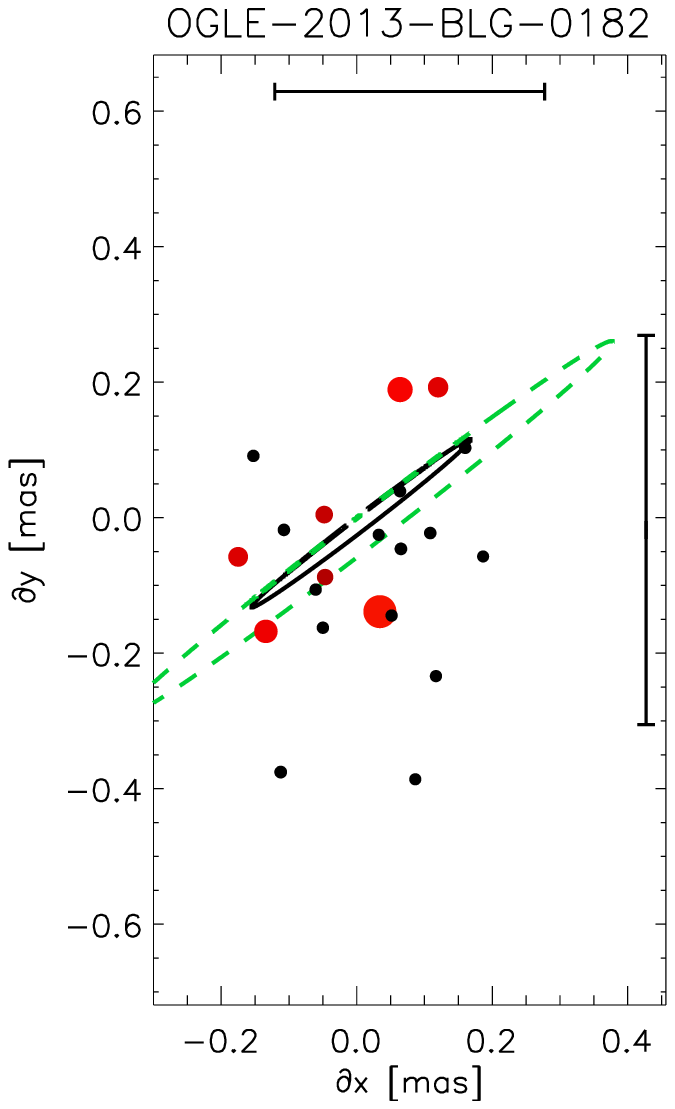}
   \hspace{0.5cm}
  \includegraphics[width=5cm, angle=0]{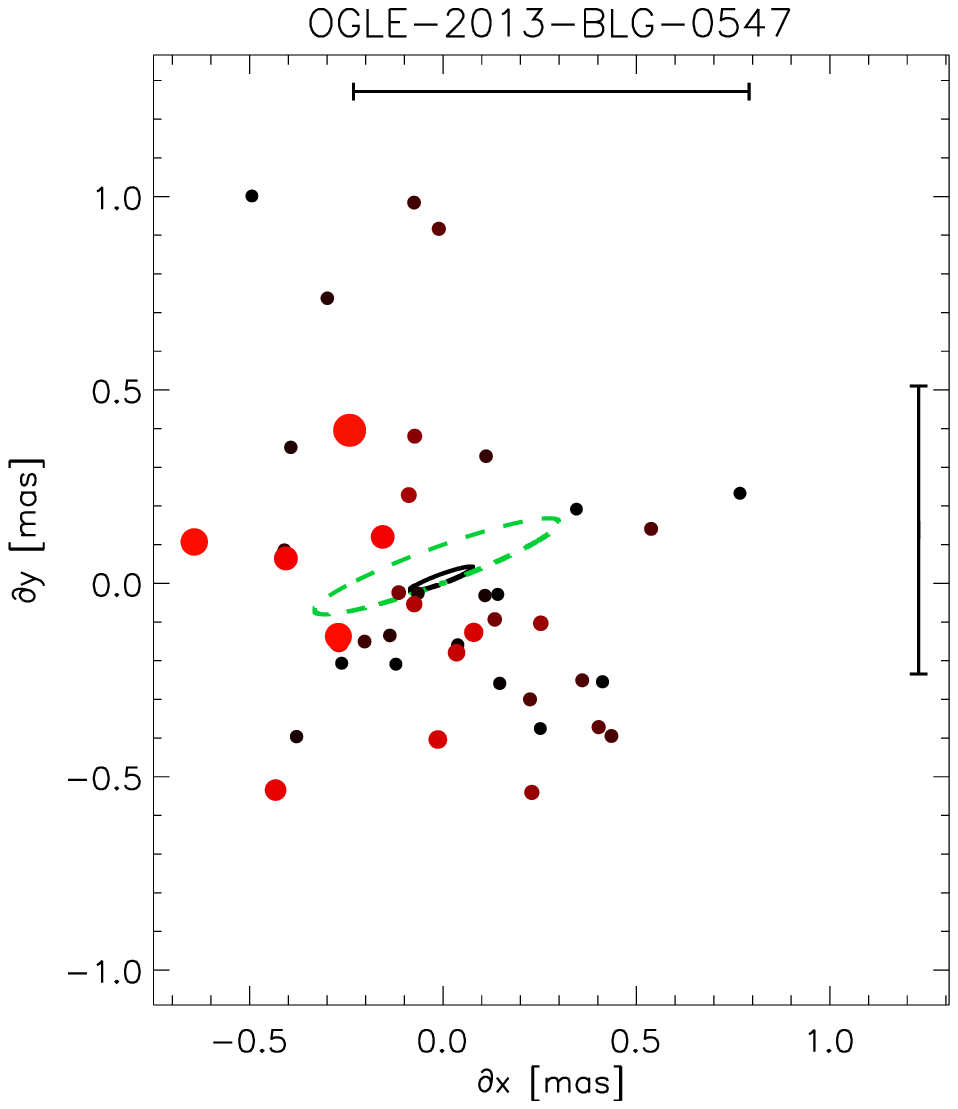}
    \vspace{0.5cm}
  \includegraphics[width=6cm, angle=0]{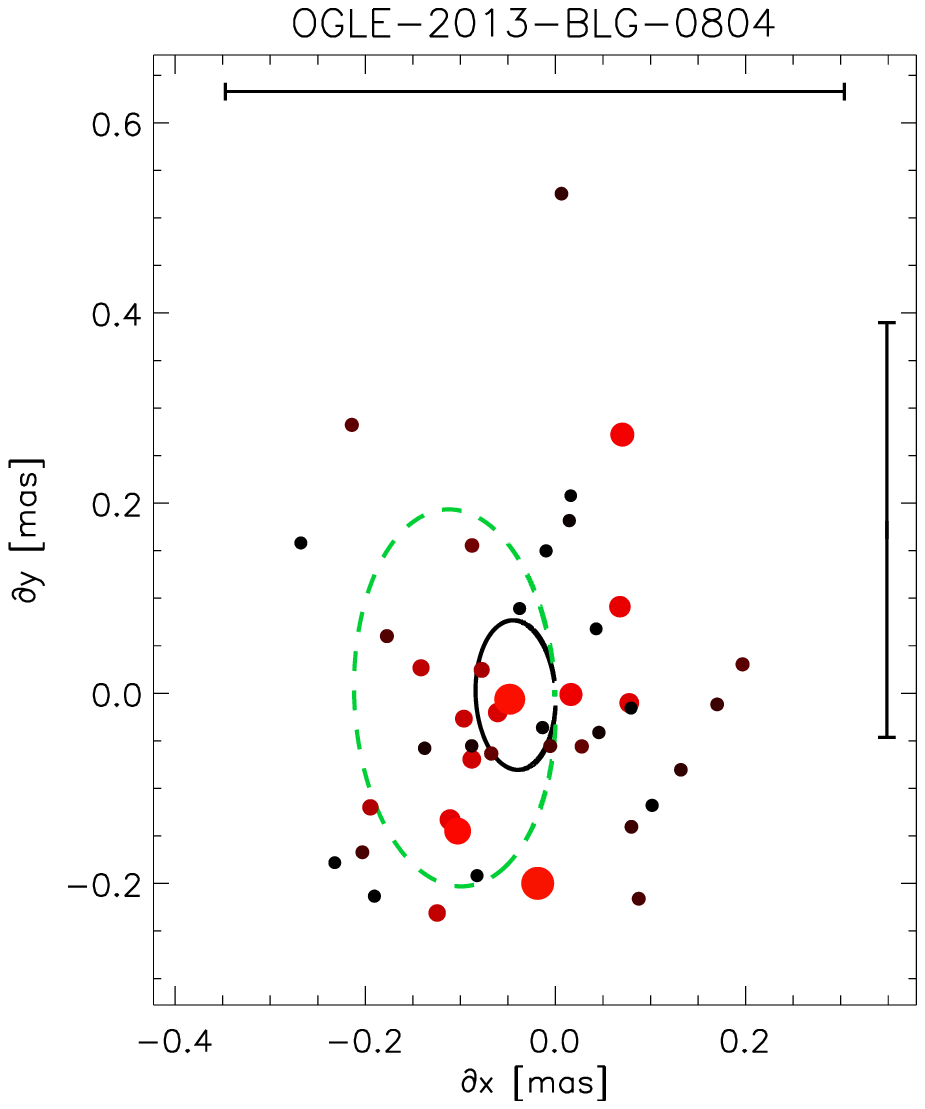}
   \hspace{0.5cm}
  \includegraphics[width=4.5cm, angle=0]{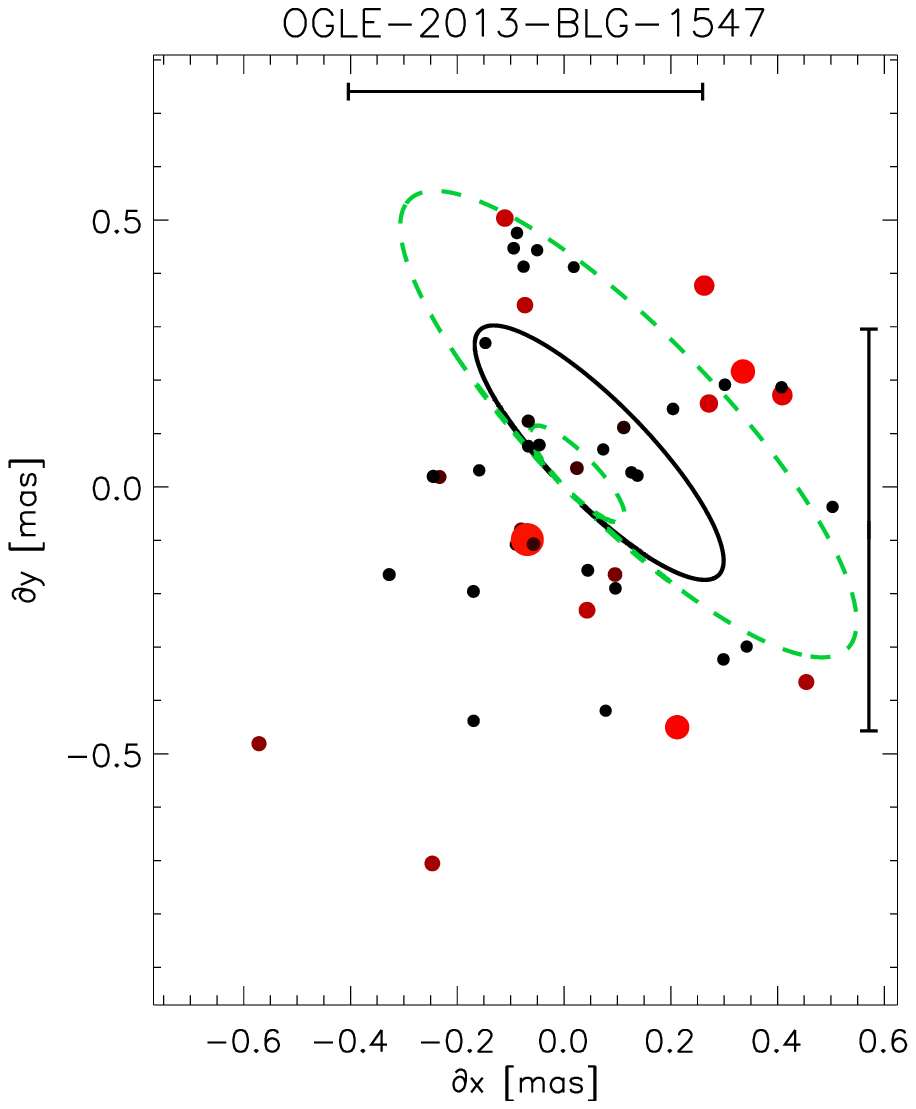}
   \hspace{0.4cm}
  \includegraphics[width=5cm, angle=0]{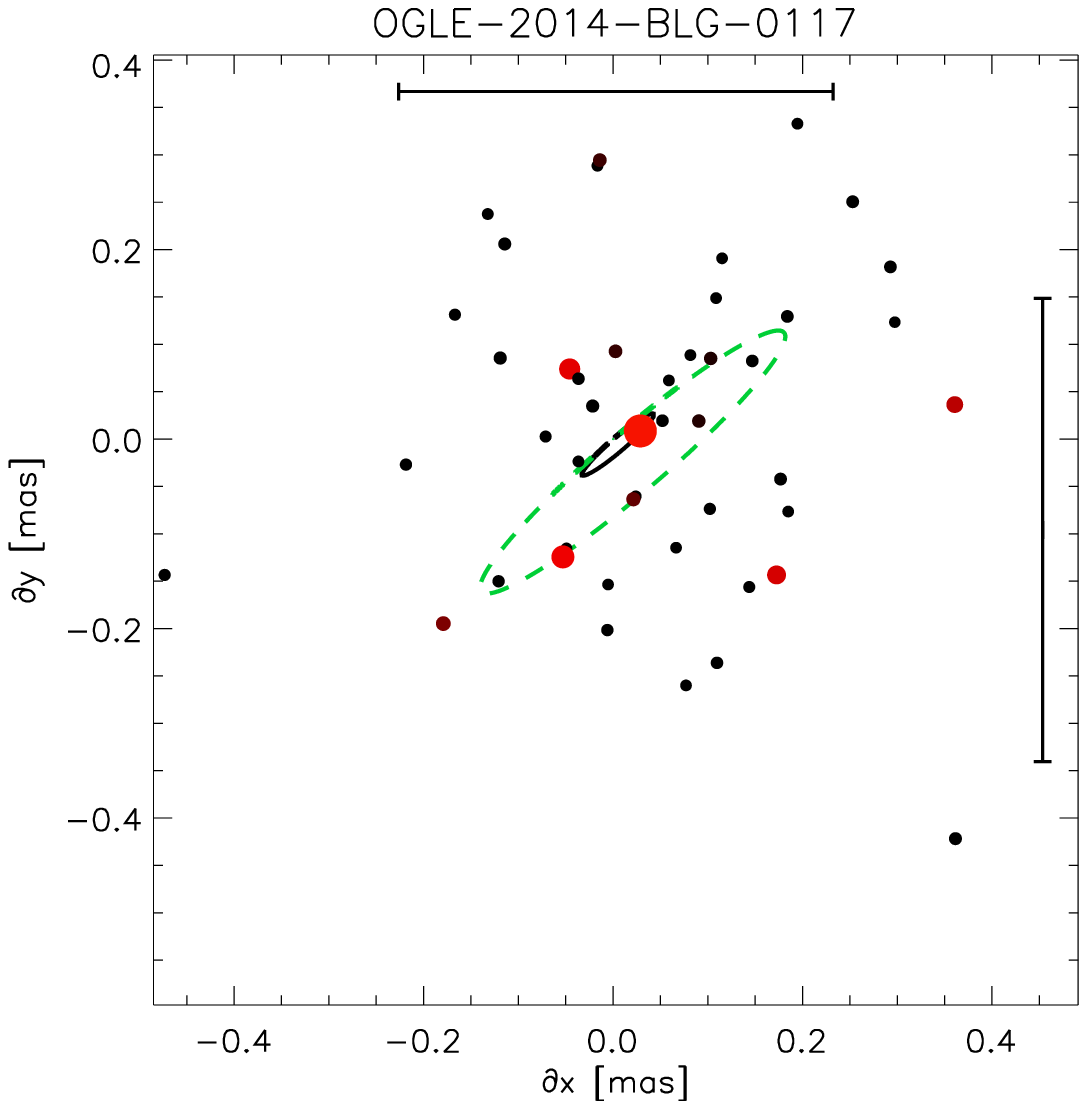}
  \includegraphics[width=12cm, angle=0]{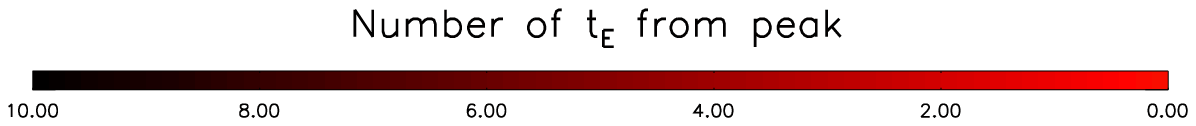}
  \caption{Same as \Fig{fig:ast1d}, but showing the residual 2-D motion of the source, after subtraction of the proper motion. Also plotted are the best-fit astrometric microlensing model (solid black ellipse), and the trajectories allowed at the edges of the 99.7\% confidence interval (green dashed ellipses). The symbol sizes and colours change according to the time of the measurement, measured in $(t-\tz)/\te$, as shown by the colour bar provided: symbols are redder and larger for measurements closer to $\tz$, and smaller and blacker further away from $\tz$. In these plots, the ($x, y$) baseline reference position of the source at $t \ll \tz$ is taken to be (0, 0). Typical error bars for the $x$ and $y$ positions are shown on the right and top edges of the plots (black lines), respectively. \label{fig:ast2d}}
\end{figure*}

The photometric and astrometric fits are shown in \Fig{fig:photfit} and Figs. \ref{fig:ast1d}-\ref{fig:ast2d}, respectively, with the parameters given in \Tab{tab:fitpar}.  

Only two events, OGLE-2013-BLG-0804 and OGLE-2013-BLG-0547, have well-constrained parallax parameters, despite the former being a low-magnification event ($\uz \sim 0.9$); in this case, the precise \hst photometry were crucial to constraining the parallax, as well as VIMOS observations.

\begin{table*}
  \vspace{0.5cm}
\begin{center}
  \begin{tabular}{cccccccccccc}
\hline
Event &$\tz$	&$\te$	&$\uz$	&$\pien$	&$\piee$	&$\alpha$	&$\thetae$&$V_{\mathrm{S}}$	&$I_{\mathrm{S}}$\\
\hline	   
OGLE-2012-BLG-0645 & 6061.976$_{-0.007}^{+0.008}$ & 7.40$_{-0.26}^{+0.26}$ & 0.087$_{-0.004}^{+0.004}$ & $-$ & $-$ & 0.95$_{-0.68}^{+0.58}$ & 0.54$_{-0.37}^{+0.73}$ &21.69 &20.28 \\
OGLE-2013-BLG-0182 & 6365.939$_{-0.007}^{+0.007}$ & 21.98$_{-0.32}^{+0.32}$ & 0.072$_{-0.002}^{+0.002}$ & $-$ & $-$ & 3.80$_{-0.48}^{+0.37}$ & 0.58$_{-0.23}^{+0.25}$ &20.52 &19.24 \\
OGLE-2013-BLG-0547 & 6418.753$_{-0.074}^{+0.073}$ & 50.62$_{-0.94}^{+0.89}$ & 0.200$_{-0.006}^{+0.006}$ & -0.85$_{-0.19}^{+0.25}$ & -0.17$_{-0.05}^{+0.06}$ & 0.28$_{-0.75}^{+0.49}$ & 0.22$_{-0.17}^{+0.25}$ &21.88 &20.69 \\
OGLE-2013-BLG-0804 & 6442.253$_{-0.041}^{+0.041}$ & 36.62$_{-0.65}^{+0.71}$ & 0.893$_{-0.029}^{+0.028}$ & -0.41$_{-0.04}^{+0.05}$ & -0.37$_{-0.04}^{+0.05}$ & 1.57$_{-0.40}^{+0.39}$ & 0.26$_{-0.13}^{+0.14}$ &20.64 &19.50 \\
OGLE-2013-BLG-1547 & 6511.909$_{-0.094}^{+0.127}$ & 25.09$_{-1.44}^{+1.39}$ & 0.412$_{-0.052}^{+0.039}$ & $-$ & $-$ & 5.51$_{-0.27}^{+0.30}$ & 0.96$_{-0.35}^{+0.21}$ &21.46 &20.33 \\
OGLE-2014-BLG-0117 & 6718.075$_{-0.019}^{+0.018}$ & 15.85$_{-0.31}^{+0.33}$ & 0.217$_{-0.007}^{+0.007}$ & $-$ & $-$ & 3.88$_{-0.54}^{+0.41}$ & 0.15$_{-0.10}^{+0.13}$ &19.78 &18.60 \\
\hline
  \end{tabular}
  \caption{Best-fit parameters for the combined fits to the photometry and astrometry for each event, with 1-$\sigma$ error bars. $V_{\mathrm{S}}$ and $I_{\mathrm{S}}$ refer to the deblended baseline magnitudes of the source, in the \hst F606W and F814W filters, respectively. \label{tab:fitpar}}
  \end{center}
\end{table*}

\section{Results and discussion}\label{sec:discussion}

The photometric models are well constrained by the photometry, particularly by our VIMOS and \hst photometry, thanks to the precision of the photometry from these data sets (Fig. \ref{fig:rms_hst_vimos}), which have scatter approximately 5 and 10 times smaller than OGLE data for a typical V$\sim$20 star. However, OGLE data are also important in constraining the baseline of the source star. The photometric parameters enable us to derive tight priors for the astrometric modelling.

In the absence of astrometric microlensing signal detections, only upper limits on the mass can be derived, corresponding to the largest value of $\thetae$ that is allowed by the data. The astrometric fit parameters in four of our six events are consistent with $\thetae=0$ at the 3-$\sigma$ level, corresponding to a no-microlensing model, which is also clear from the plots of astrometric fits in Figs. \ref{fig:ast1d}-\ref{fig:ast2d}, meaning that no lower limit can be derived for the lens mass. Because we know from photometry that microlensing did occur, the posterior distributions from the MCMC astrometric fit allow us to place limits on the size of $\thetae$, and therefore on lens masses, when combined to distance constraints. The mass limits derived from these constraints are given in \Tab{tab:masslimits}. For two other events, the 3-$\sigma$ lower limit on $\thetae$ is larger than zero, but the lack of parallax measurements means that this does not translate into significant lower lens mass limits. 

\begin{figure*}
  \centering
  \includegraphics[width=8cm, angle=0]{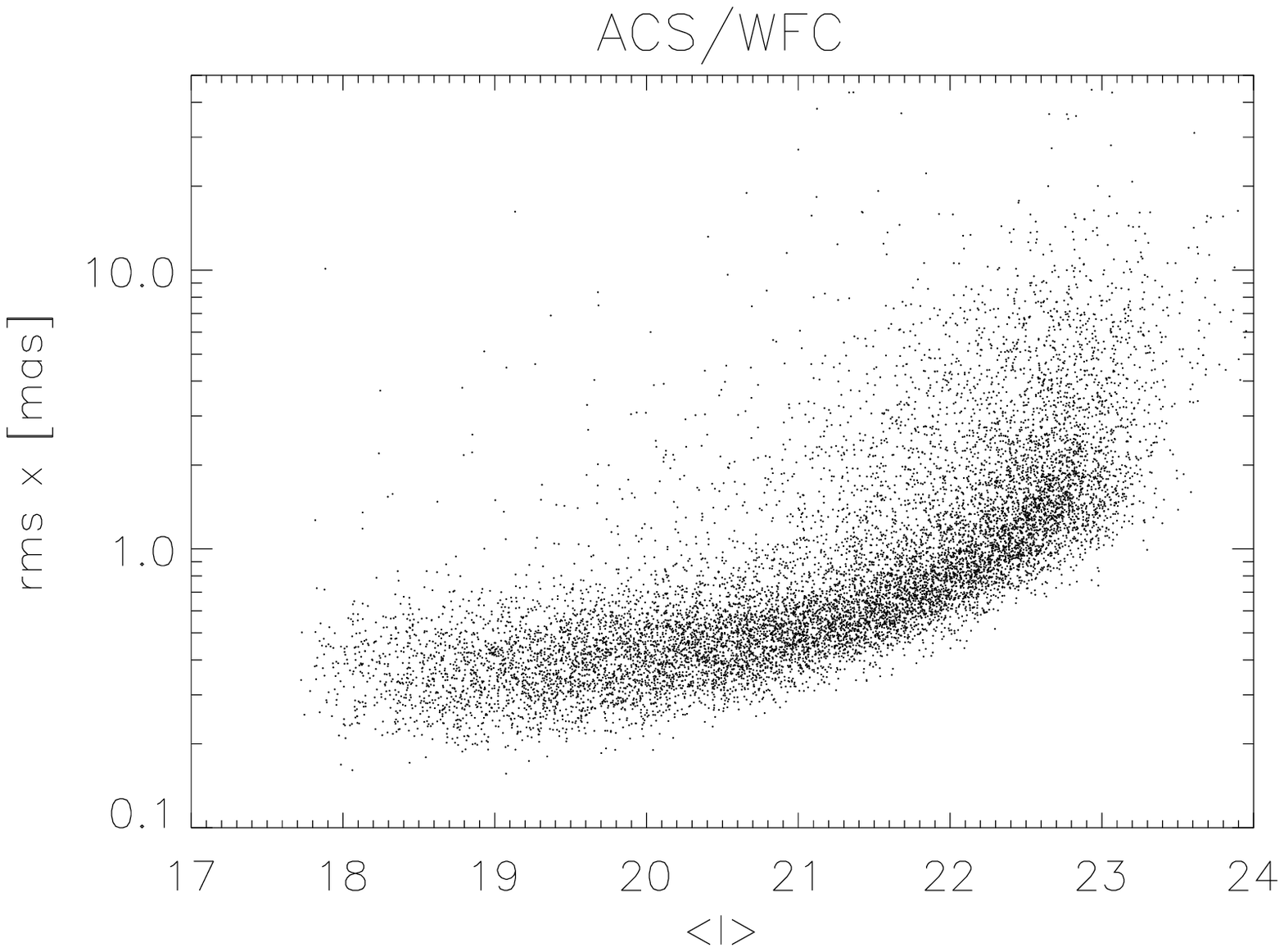}
  \includegraphics[width=8cm, angle=0]{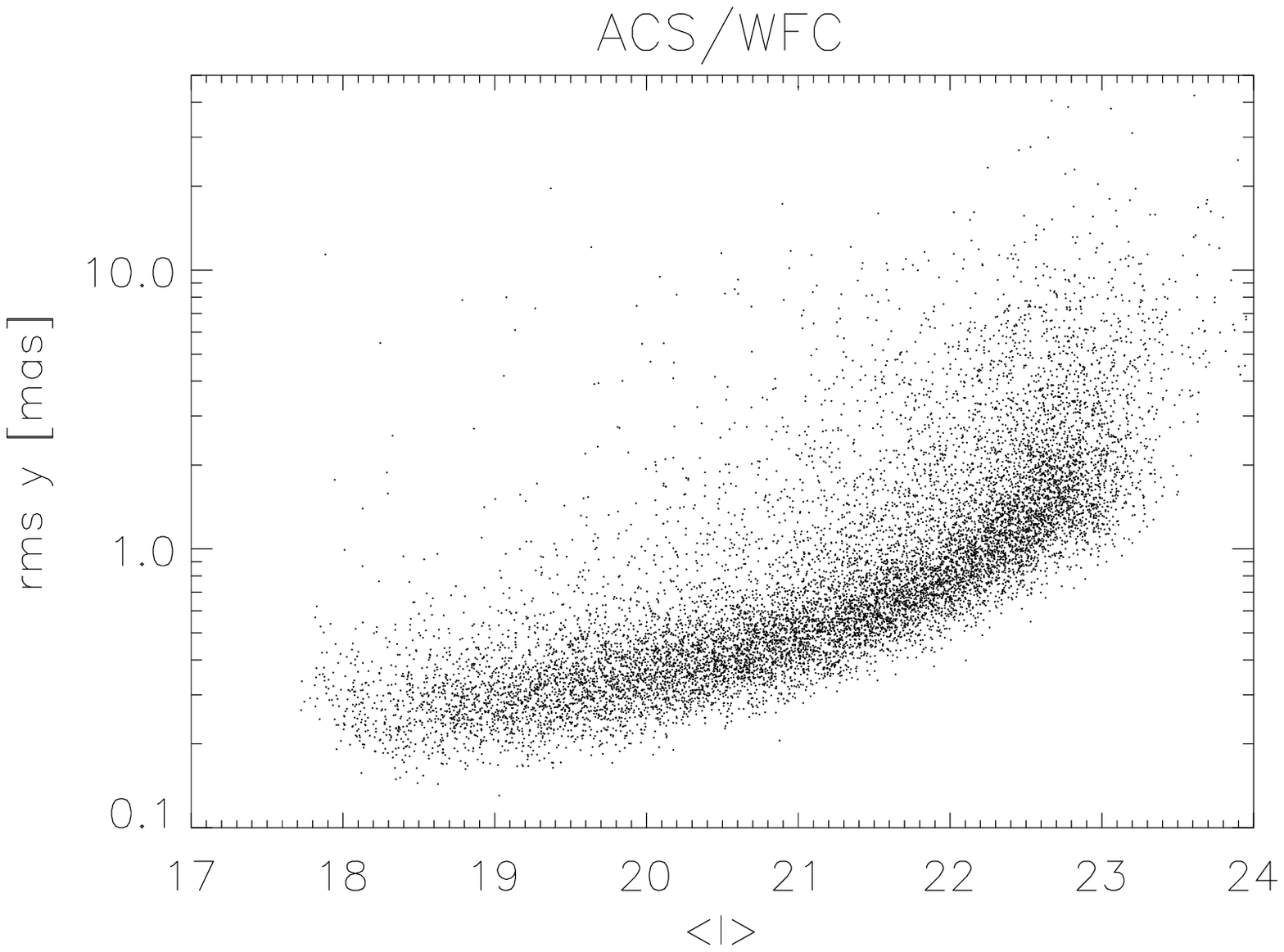}
  \includegraphics[width=8cm, angle=0]{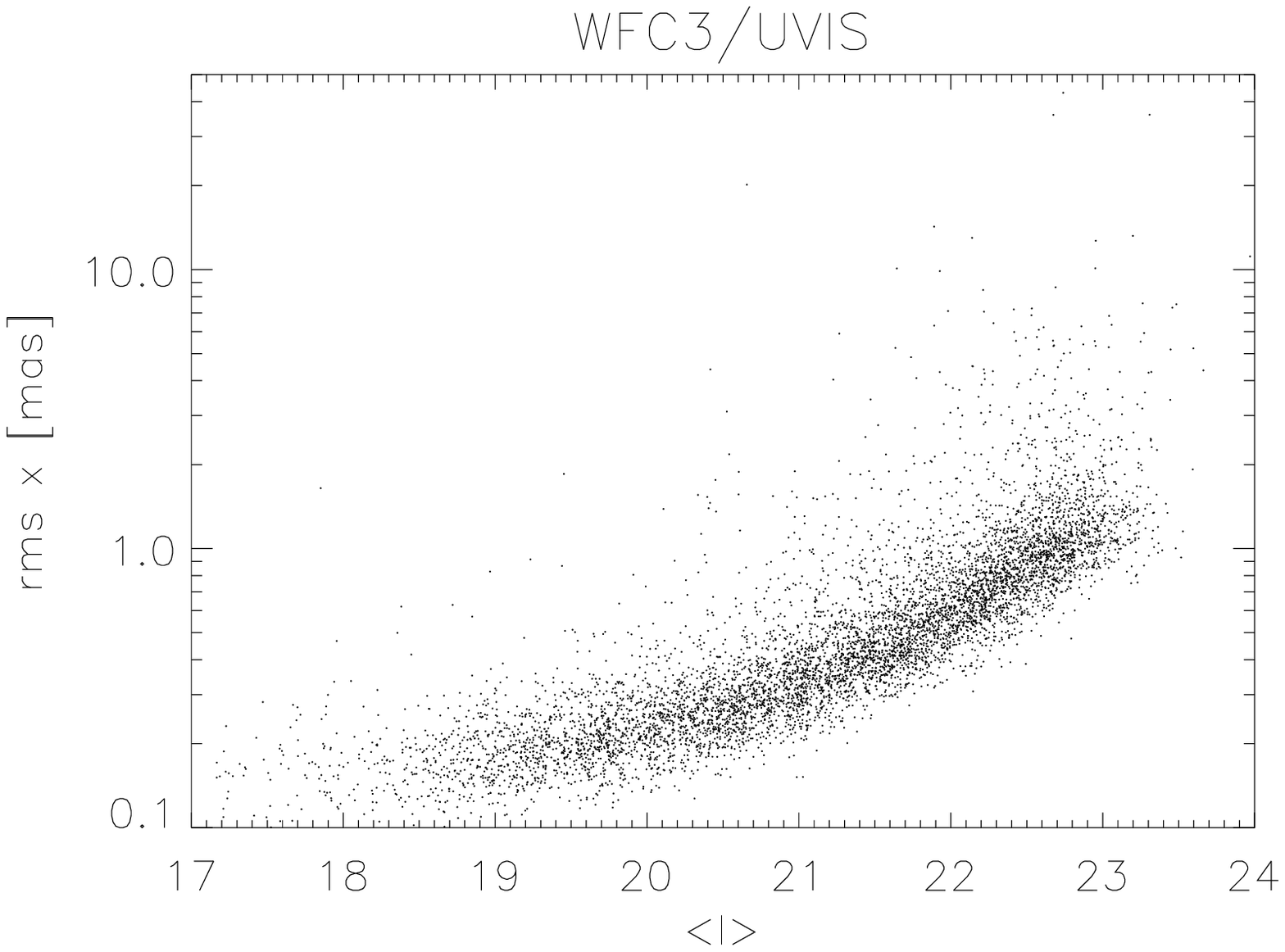}
  \includegraphics[width=8cm, angle=0]{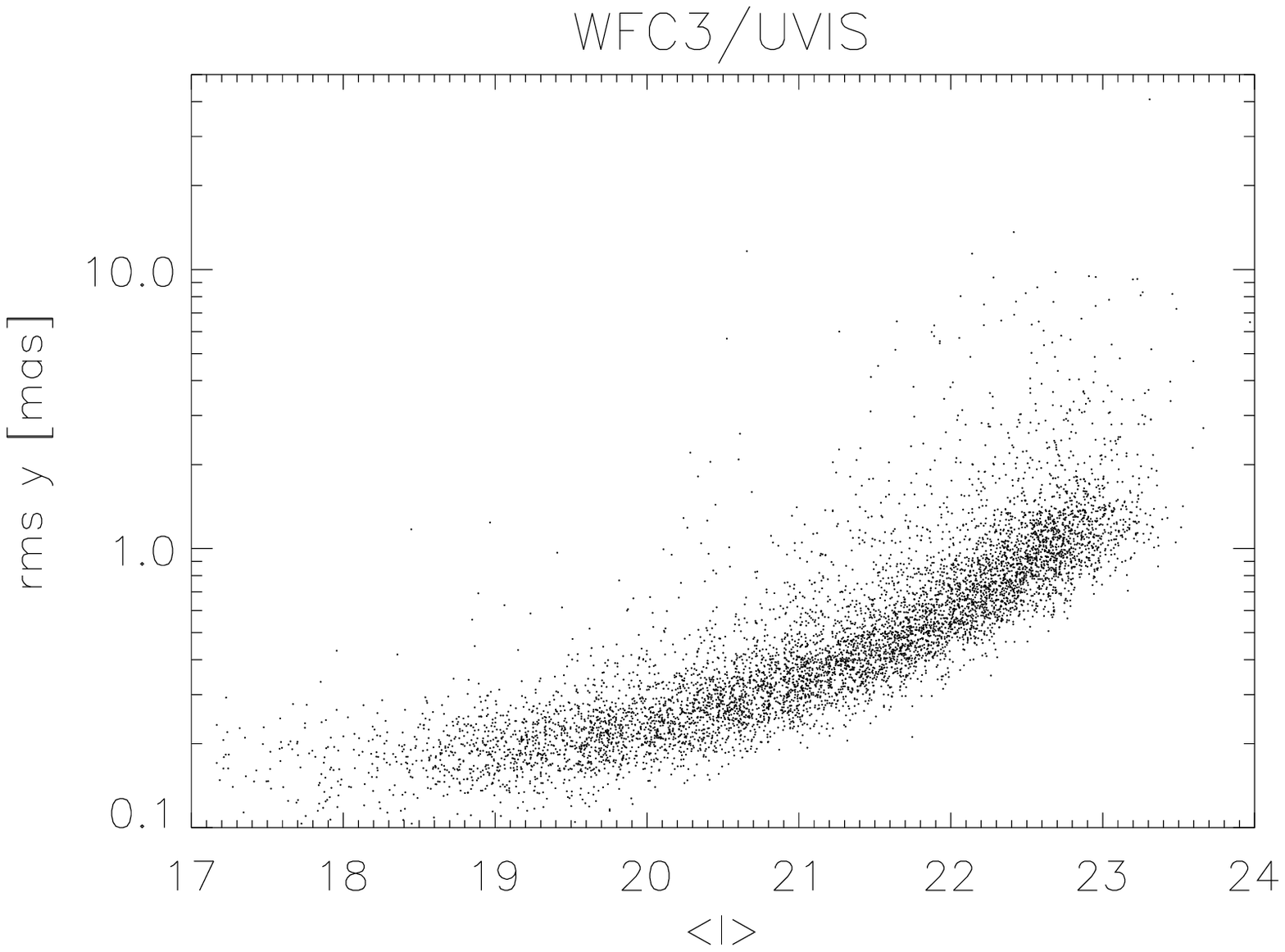}
  \caption{The rms of astrometric curves in the $x$ (left) and $y$ (right) position measurements for ACS (top) and WFC3 (bottom) data, plotted against the mean $I$ magnitude, for a representative sample of stars in our \hst observations. \label{fig:astrms}}
\end{figure*}

The astrometric models are well constrained, thanks to the astrometric precision of our measurements, which is consistent with the expected astrometric precision $z_i$ at the $i^{\mathrm{th}}$ epoch, expressed by \cite{kuijken02}, 

\begin{equation}\label{eq:precision}
z_i=0.7\frac{\mathrm{FWHM}}{\mathrm{SNR}\times \sqrt{N_i}}\, ,
\end{equation}

\noindent
where FWHM is the full-width half-maximum of the star's PSF, and $N_i$ is the number of images at the $i^{\mathrm{th}}$ epoch. Therefore, the constraints on $\thetae$ that we obtain from the astrometric measurements are as good as what can be expected for the depth and number of observations in our program. \Fig{fig:astrms} shows the rms scatter of astrometric light curves for a representative sample of stars in our observations.

\begin{figure*}
  \centering
  \includegraphics[width=8cm, angle=0]{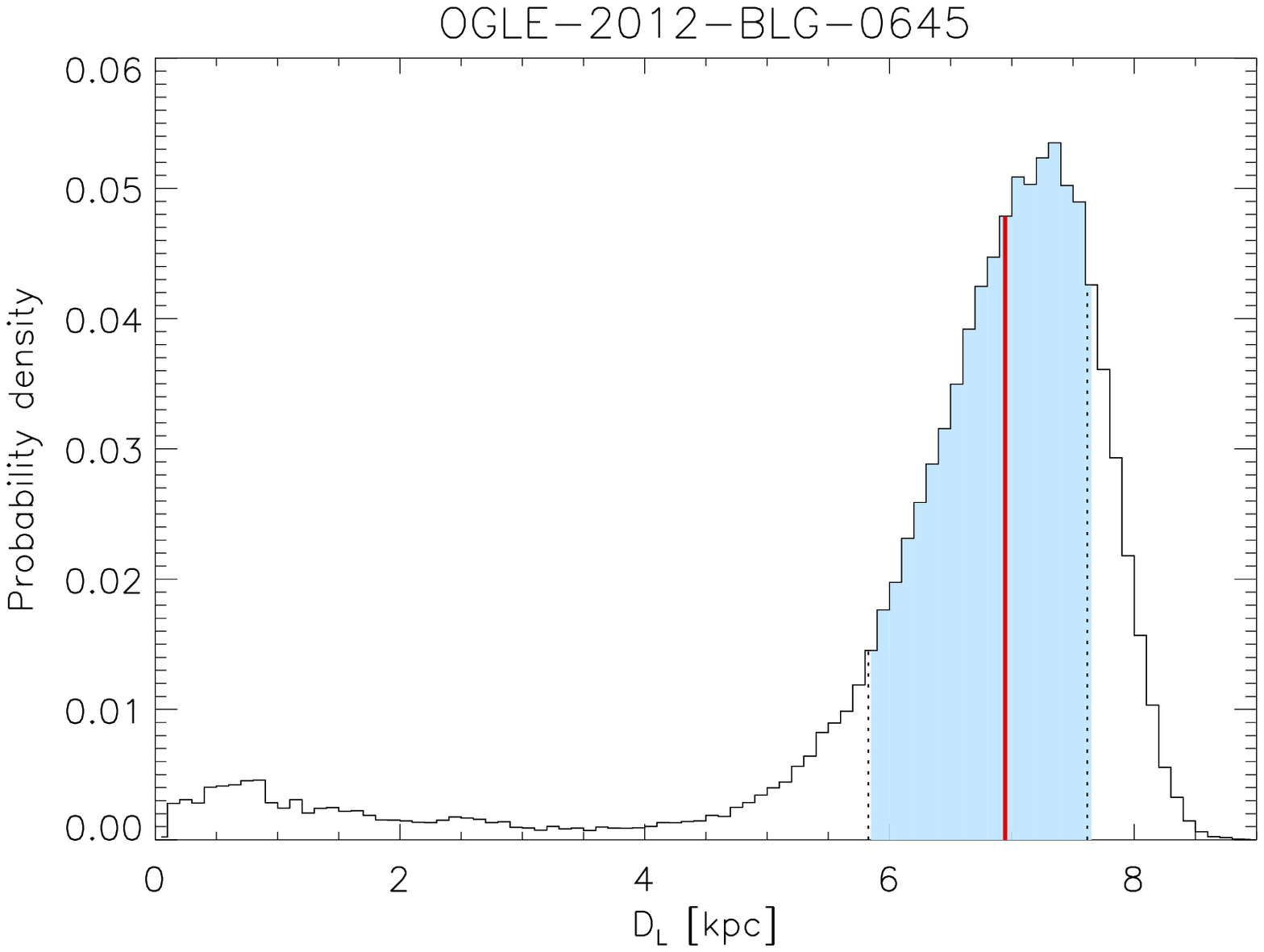}
  \includegraphics[width=8cm, angle=0]{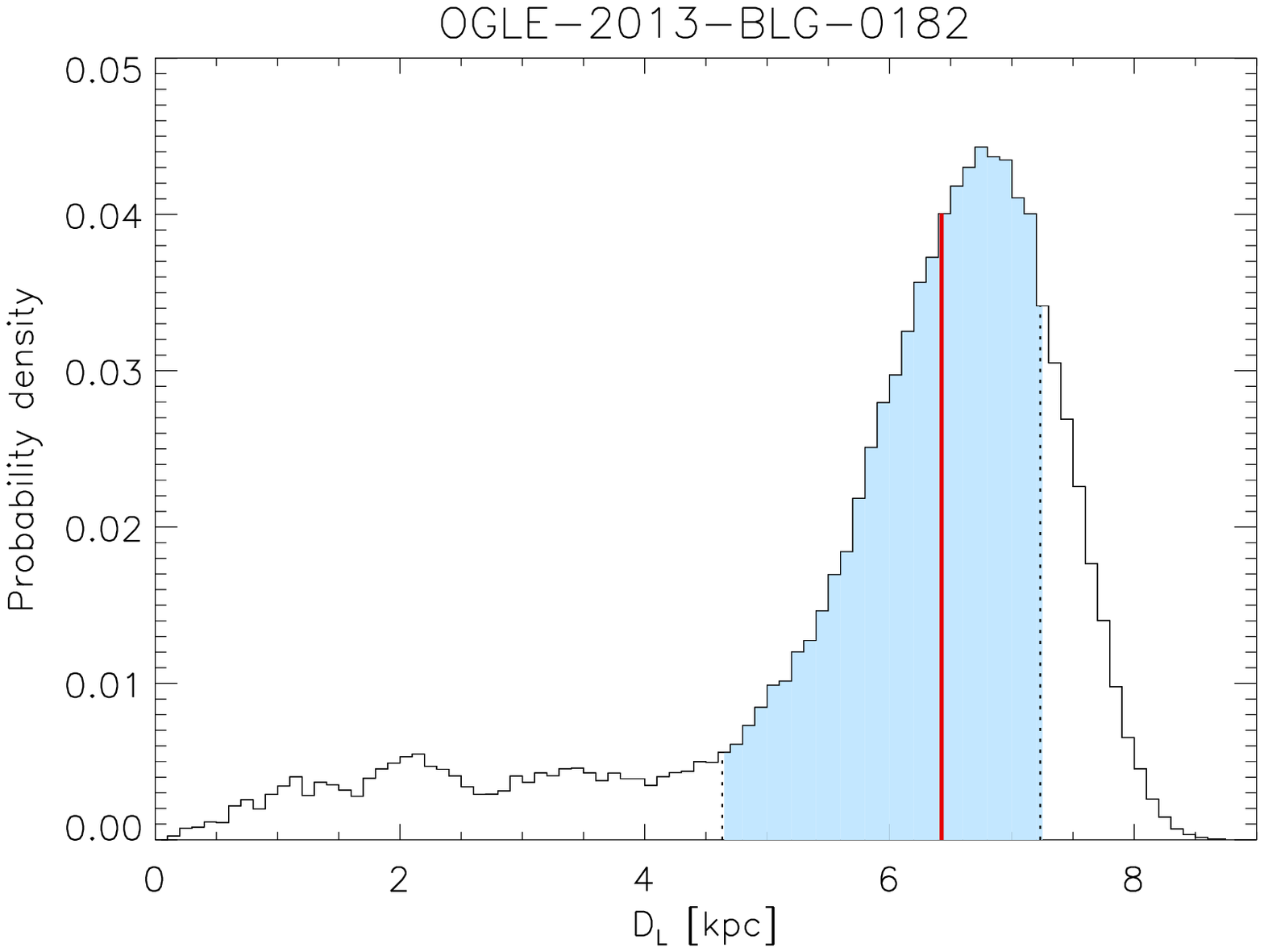}
  \includegraphics[width=8cm, angle=0]{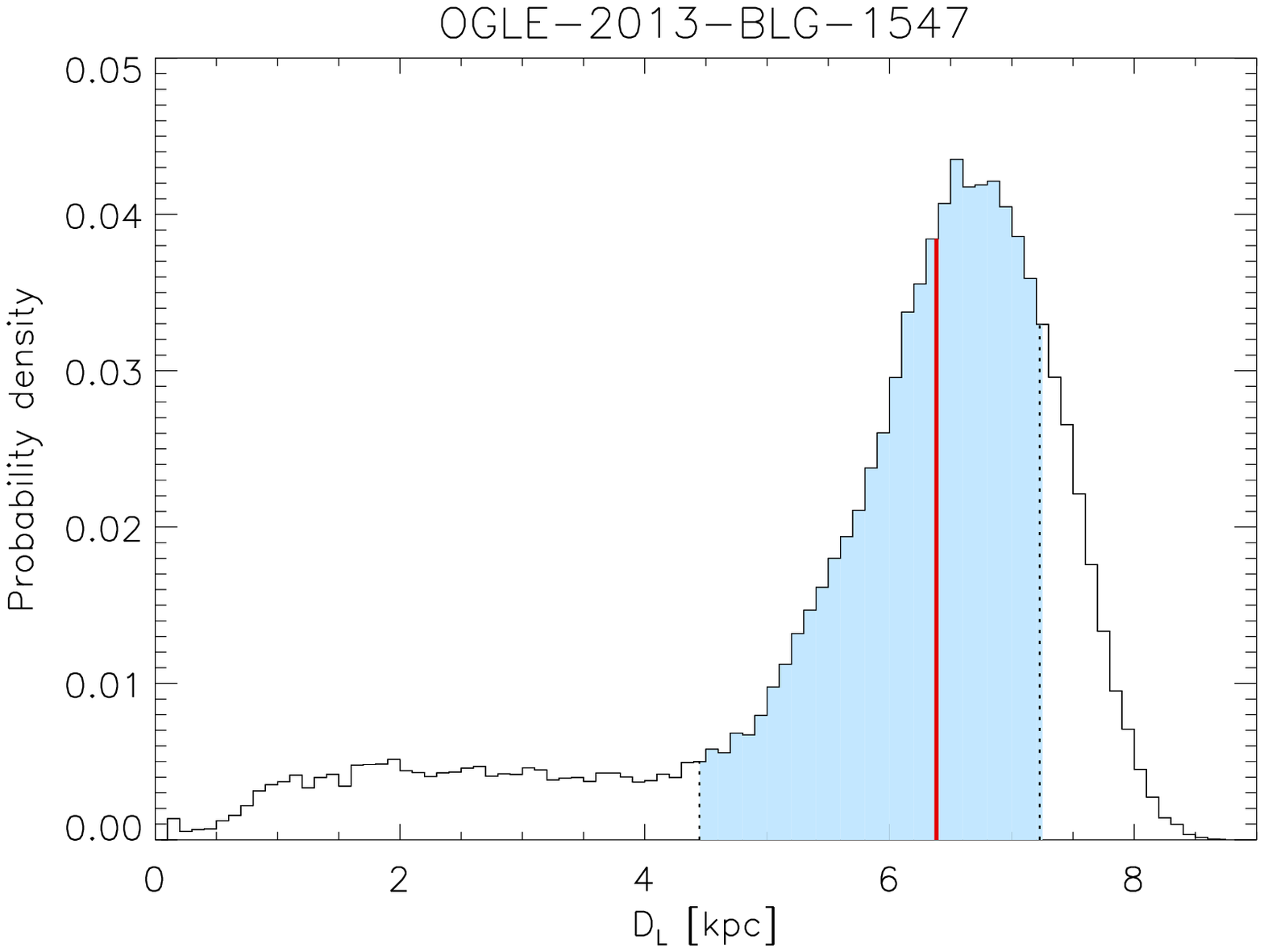}
  \includegraphics[width=8cm, angle=0]{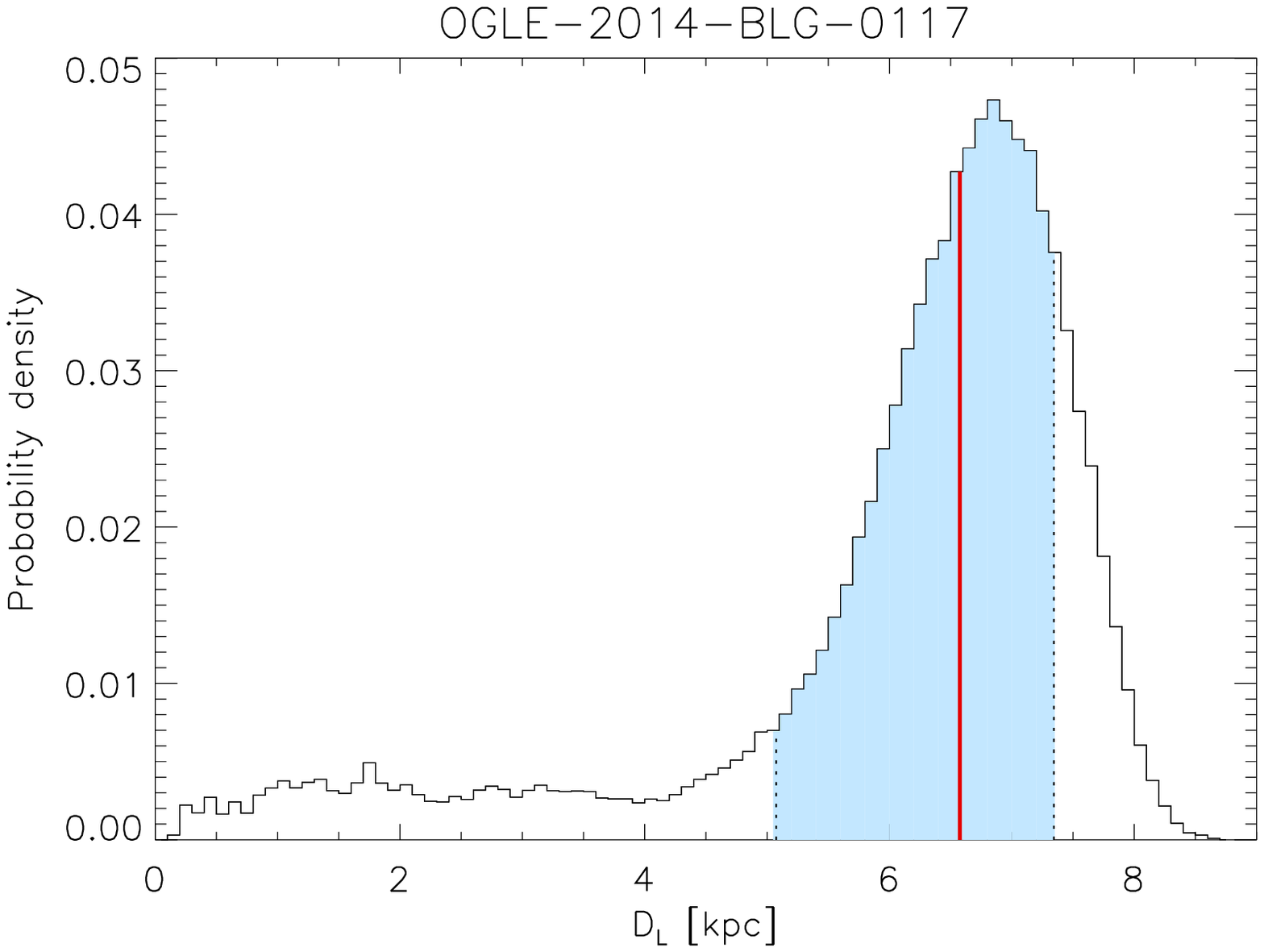}
\caption{The distribution of the lens distance $\dl$ for the four events for which parallax is not constrained by the light curves. These distributions were obtained using the algorithm described in \cite{tsapras16}, using a source distance of $8.0 \pm 0.3$ kpc. The median value of $\dl$ is shown as a vertical solid red line, and the 68.3\% confidence interval is shown as the shaded blue region, delimited by vertical dotted lines. \label{fig:dldist}}
\end{figure*}

The effect of parallax is constrained in just two of the six events, meaning that the distance to the lens $D_L$ is not constrained for the other four; for these, we must therefore use probabilistic distances derived from each event's fitted timescale $\te$ and Galactic models. Here we use the algorithm detailed in \cite{tsapras16}, which is itself based on the approach of \cite{dominik06}; resulting distributions of $\dl$ for these four events are shown in \Fig{fig:dldist}. Using these probabilistic distances leads to essentially unconstrained lens masses, due to the large spread of allowed values of $\dl$; this is reflected in large error bars and very large mass upper limits.

\subsection{Individual events}
\subsubsection*{OGLE-2012-BLG-0645}
This event is short, with $\te=7.4\pm0.3$ days, meaning that parallax is not constrained. \hst data only covers the event before and after the peak, at small magnification, but the small value of $\uz$ means that the OGLE data are sufficient to constrain the photometric parameters well. The astrometric model is consistent with $\thetae=0$ at 3 $\sigma$, with a 99.7\% confidence interval of $\thetae=[0, 3.18]$ mas, but is poorly constrained due to large $x$ and $y$ astrometric scatter of $1.26$ and $0.92$ mas, respectively. Combined with the statistical distance of $7.0_{-1.1}^{+0.7}$kpc, the astrometric parameters do not allow us to derive a meaningful upper mass limit for the lens; nevertheless, the formal mass limit estimated for is quoted in \Tab{tab:masslimits} for all events.

\subsubsection*{OGLE-2013-BLG-0182}

The photometric parameters for this event are well constrained, because it is relatively bright with a deblended source magnitude $I_S=19.2$ mag, and the impact parameter is small, $\uz=0.07$, but the parallax is not measured. The astrometric model is tightly constrained by the WFC3 observations, with a scatter of 0.39 mas in the $x$ position and 0.53 mas in the $y$ position. The resulting 99.7\% confidence interval for $\thetae$ of $[0.01, 1.43]$ mas, but no meaningful mass limits are derived due to the lack of parallax constraints.

\subsubsection*{OGLE-2013-BLG-0547}

This event has the longest timescale in our sample, at $\te=50$ d, which, combined with $\uz=0.2$, allowed us to constrain the parallax parameters from the light curve. The scatter in the astrometric measurements is 1.56 and 0.84 mas for the $x$ and $y$ positions, respectively, and the astrometric model yields edges of the 99.7\% confidence interval at $\thetae=[0, 1.02]$ mas; combining the various parameters gives a distance of $3.0\pm0.6$kpc, and an upper limit for the lens mass of $0.66 \msun$. 

\subsubsection*{OGLE-2013-BLG-0804}

The timescale $\te=37$ days also allowed us to constrain the parallax parameters for this event, despite a large impact parameter $\uz=0.9$. The rms scatter of the astrometric measurements is 0.40 and 0.34 mas in the $x$ and $y$ directions, respectively, allowing for good constraints on the astrometric model, with a 99.7\% confidence interval of $\thetae=[0, 0.48]$ mas. Combined with the parallax parameters, this yields a distance of $3.7\pm0.3$kpc, and an upper mass limit of 0.43 $\msun$.

\subsubsection*{OGLE-2013-BLG-1547}

The parallax is not measured for this event, and the large uncertainties in OGLE photometry, sparse \hst coverage of the light curve, and lack of a VIMOS light curve, means that the error bars on $\te$ are large. However, the astrometric model is constrained and is consistent with $\thetae > 0$ at the 3-$\sigma$ level. We find a 99.7\% confidence interval for $\thetae$ of $[0.19, 1.80]$. The rms scatter of the astrometric measurements is 0.59 and 0.85 mas in the $x$ and $y$ directions, respectively. Unfortunately, the lack of parallax measurement means that this cannot be translated directly into meaningful mass limits.

\subsubsection*{OGLE-2014-BLG-0117}

This is the brightest of the six events, and the good OGLE photometry, combined with \hst observations, allows us to constrain the photometric parameters well, but not the parallax, which means that no meaningful mass limits can be placed. The astrometric model yields a 99.7\% confidence interval for $\thetae$ of $[0, 0.58]$ mas, thanks to an astrometric scatter of 0.34 and 0.31 mas in the $x$ and $y$ astrometric measurements, respectively.

\begin{table}
  \vspace{0.5cm}
\begin{center}
  \begin{tabular}{cccccccccc}
\hline
Event &$\dl$ [kpc]	&$M_L/\msun$ &$M_{L, \mathrm{up}}/\msun$\\
\hline	   
OGLE-2012-BLG-0645 &$7.0_{-1.1}^{+0.7}$ &$1.76_{-1.66}^{+12.47}$ &403.35 \\
OGLE-2013-BLG-0182 &$6.4_{-1.8}^{+0.8}$ &$1.25_{-1.02}^{+2.70}$ &47.43 \\
OGLE-2013-BLG-0547 &$3.0_{-0.6}^{+0.6}$ &$0.03_{-0.03}^{+0.10}$ &0.66 \\
OGLE-2013-BLG-0804 &$3.7_{-0.3}^{+0.3}$ &$0.06_{-0.04}^{+0.08}$ &0.43 \\
OGLE-2013-BLG-1547 &$6.4_{-1.9}^{+0.8}$ &$3.59_{-2.64}^{+5.37}$ &87.12 \\
OGLE-2014-BLG-0117 &$6.6_{-1.5}^{+0.8}$ &$0.08_{-0.07}^{+0.35}$ &6.88 \\
\hline
  \end{tabular}
  \caption{Lens distances and masses derived from the model fits, and upper limit (99.7\% confidence interval) for the lens masses. The distance to the source is taken to be $\ds=8.0\pm0.3$ kpc. The lens masses in column 3 are given with 1-$\sigma$ percentile error bars. \label{tab:masslimits}}
  \end{center}
\end{table}

\subsection{Future prospects for astrometric measurements}

We are currently limited by ground-based photometry to measure the microlensing parallax in large-scale surveys. Although Gaia is a large space-based astrometric mission, it cannot reach the required precision toward the crowded fields where microlensing is more likely to be detected. In the future, it will be possible to measure parallax routinely from space, without the need for parallel ground-based monitoring, which in any case will struggle to reach the required photometric precision for the fainter sources observed from space. This will be done either with high-precision, dense sampling of photometric light curves to detect the effect of parallax discussed in \Sec{sec:microlensing}, or via measurement of the so-called microlensing ``space parallax". The former is more easily detected in long ($\te\gtrsim40$ days) events, but we carried out some rough simulations of deviations from the PSPL model caused by the effect of parallax by various event configurations. We found that in some cases, the photometric precision afforded by space-based observations could allow us to detect the effect of parallax in some events as short as $\tE \sim 15$ days, with observations from a single space-based observatory. Space parallax, on the other hand, manifests itself as a time shift between a microlensing light curve observed from Earth and observations from space due to the shift in perspective caused by the distance between Earth and the space-based observatory. This effect has already been used to constrain the distance to lens systems in a few cases using the \textit{Spitzer Space Telescope} \citep[e.g.][]{udalski15, yee15}. Although such measurements usually require a large distance of at least a few thousandths of an AU between Earth and the space-based observatory, it will be possible to constrain the parallax in this way using the \textit{Wide-Field Infrared Survey Telescope} (\textit{WFIRST}), which will be orbiting with an apoapsis of $\sim0.005$ AU. Furthermore, the planned cadence of observations for the \textit{WFIRST} Microlensing Survey \citep{spergel15} will be sufficient to constrain the microlensing parallax in many of the longer microlensing events from the light curve alone \citep{yee13}. Additional observations from the ground, for example through the use of target-of-opportunity (ToO) observations, will also provide further opportunities to measure the effect of parallax when this cannot be achieved from the \textit{WFIRST} light curve alone. Combined with astrometric measurements similar to those presented in this paper, this will yield routine mass measurements for many lenses, enabling us to constrain the properties of many of the detected systems, including exoplanets, single stars, and compact objects.

\section{Conclusions}\label{sec:conclusion}

In this paper, we have demonstrated that routine measurements of the size of the Einstein ring radius can be achieved using astrometric microlensing, and that this will be a powerful way to constrain the mass of objects detected by space-based microlensing surveys. Currently, only the small fraction of events in which second-order effects are detected have constraints on $\thetae$, which then becomes a limiting factor in determining lens masses when source stars are faint and have a small angular size. Since most source stars in microlensing events are M dwarfs, this means that most microlenses cannot have their masses constrained strongly. On the other hand, the methods we developed for this study can be applied to all stars for which precise enough astrometry can be measured, and will be applied to a large-scale search for stellar-mass black holes, using the same data set. 

In the future, the astrometric precision afforded by the \textit{James Webb Space Telescope} (\textit{JWST}), and particularly by the \textit{WFIRST} Microlensing Survey \citep{spergel15}, down to $\sim50\,\muas$ per measurement for stars with $V\sim23$, will mean that exoplanet and black hole masses will be much better constrained on a routine basis. This is a crucial important step towards deriving large-sample demographics from future microlensing observations.\\

\section*{Acknowledgements}

NK acknowledges support from \hst grants GO-12586, GO-13057, GO-13463 (PI: Sahu) and AR-14571 (PI: Kains). The OGLE project has received funding from the National Science Centre, Poland, grant MAESTRO 2014/14/A/ST9/00121 to AU. MZ acknowledges support by the Ministry of Economy, Development, and Tourism's  
Millennium Science Initiative through grant IC120009, awarded to The Millennium Institute of Astrophysics (MAS), by Fondecyt Regular 1150345 and by the BASAL-CATA Center for Astrophysics and Associated Technologies PFB-06.

\software{\texttt{hst2xym} \citep{anderson06}, \texttt{KS2} \citep[e.g.][]{anderson00}, \texttt{DanDIA} \citep{bramich13, bramich08}, OGLE pipeline \citep{udalski03}}

\bibliographystyle{aasjournal}
\bibliography{../thesisbib}

\label{lastpage}

\end{document}